\documentclass[11pt]{iopart}
\usepackage{cite}
\usepackage{epsfig}
\usepackage{amssymb}
\usepackage{graphicx}
\usepackage{epsf}
\usepackage{geometry}
\usepackage[dvips]{color}

\begin{document}

\newcommand{\deriv}[2]{{\frac{d#1}{d#2}}}
\newcommand{\be}{\begin{equation}}
\newcommand{\ee}{\end{equation}}
\newcommand{\bea}{\begin{eqnarray}}
\newcommand{\eea}{\end{eqnarray}}
\newcommand{\rr}{{\bf r}}
\newcommand{\rt}{\rr^\perp}
\newcommand{\kk}{{\bf k}}
\newcommand{\kkt}{\kk^\perp}
\newcommand{\p}{{\bf p}}
\newcommand{\q}{{\bf q}}
\newcommand{\qt}{\q^\perp}
\newcommand{\X}{({\bf x})}
\newcommand{\Y}{({\bf y})}
\newcommand{\x}{{\bf x}}
\newcommand{\ff}{{\bf f}}
\newcommand{\uu}{{\bf u}}
\newcommand{\xx}{{\bf x}}
\newcommand{\EE}{{\bf E}}
\newcommand{\VV}{{\bf V}}
\newcommand{\y}{{\bf y}}
\newcommand{\U}{{\bf u}}
\newcommand{\w}{{\bf \omega}}
\newcommand{\D}{{\bf \nabla}}
\newcommand{\W}{\omega_\kk}
\newcommand{\za}{\alpha}
\newcommand{\zb}{\beta}
\newcommand{\zd}{\delta}
\newcommand{\ze}{\epsilon}
\newcommand{\zg}{\gamma}
\newcommand{\zl}{\lambda}
\newcommand{\zL}{\Lambda}
\newcommand{\zp}{\pi}
\newcommand{\zs}{\sigma}
\newcommand{\zt}{\tau}
\newcommand{\zE}{I\hskip-3.7pt E}
\newcommand{\zF}{\Phi}
\newcommand{\zN}{I\hskip-3.4pt N}
\newcommand{\zR}{I\hskip-3.4pt R}
\newcommand{\zw}{\omega}
\newcommand{\zW}{\Omega}
\newcommand{\zC}{{\mathbb C}}
\newcommand{\zD}{{\Delta}}
\newcommand{\zG}{{\Gamma}}
\newcommand{\veps}{\varepsilon}
\newcommand{\OM}{({\bf ***\ldots***})}
\newcommand{\EM}{({\bf $\leftarrow$***})}
\newcommand{\BM}{({\bf ***$\rightarrow$})}
\newcommand{\WFR}{$\Omega$-FR }
\newcommand{\WFRs}{$\Omega$-FRs }
\newcommand{\LFR}{$\Lambda$-FR }
\newcommand{\Wt}{\overline{\Omega}_{t_0,t_0+\tau}}
\newcommand{\Ft}{\overline{\mathcal{O}}_{t_0,t_0+\tau}}
\newcommand{\Fz}{\overline{\mathcal{O}}_{0,\tau}}
\newcommand{\FSS}{\mathcal{O}_{\tau}}
\newcommand{\Wz}{\overline{\Omega}_{0,\tau}}
\newcommand{\WSS}{\Omega_{\tau}}
\newcommand{\LSS}{\Lambda_{\tau}}

\newcommand{\noi}{\noindent}

\newcommand {\bdm} {\begin{displaymath}}
\newcommand {\edm} {\end{displaymath}}
\newcommand {\ba}  {\begin{array}}
\newcommand {\ea}  {\end{array}}
\newcommand {\mapx} {\Phi_{\epsilon}}
\newcommand {\Hx}    {{\mathcal H}_{\hat x}}
\newcommand {\muje} {\mu_j(\eps)}
\newcommand {\ugns}[1] {{\bf u}^{gns}(#1)}
\newcommand {\ugnsi}[1] {{\bf u}^{gns}_i(#1)}
\newcommand {\uns}[1] {{\bf u}^{ns}(#1)}
\newcommand {\parno} {\par\noindent}
\newcommand {\para} [1]{ \par\noindent {\bf #1} \par\noindent}
\newcommand {\gam} [1]{\gamma_{{\vec k}_{#1}}}
\newcommand {\norgq} [1] {| \gamma_{{\vec k}_{#1}} |^2}

\newcommand{\td}{T_r}
\newcommand{\ts}{T_\ell}
\newcommand{\xt}{X_t}
\newcommand{\ft}{f_t}
\newcommand{\CFt}{{\cal F}_t}
\newcommand{\vt}{{\cal V}_t}
\newcommand{\kt}{{\cal K}_t}
\newcommand{\rot}{\rho_t}
\newcommand{\cL}{{\tt L}}
\newcommand{\cT}{{\cal T}^{(N)}}
\newcommand{\kT}{{\cal T}}
\newcommand{\cTa}{{\cal T}^\ast}
\newcommand{\se}{{(s)}}
\newcommand{\zP}{\mbox{P}}
\newcommand{\zM}{\mbox {M}}
\newcommand{\pp}{p^+}
\newcommand{\cmin}{\underline{\cal C}}
\newcommand{\cmed}{{\cal C}_{av}}
\newcommand{\cmax}{\bar{\cal C}}
\newcommand{\ie}{i.e.\ }



\title[Fluctuation Theorems]{Fluctuations in Nonequilibrium
Statistical Mechanics: Models, Mathematical Theory, Physical Mechanisms}

\author{Lamberto Rondoni, Carlos Mej\'{\i}a-Monasterio}
\address{Dipartimento di Matematica, Politecnico di Torino,
Corso Duca degli Abruzzi 24 I-10129 Torino, Italy }
\eads{\mailto{lamberto.rondoni@polito.it}, \mailto{mejia@calvino.polito.it}}
\begin{abstract}
  The fluctuations in nonequilibrium systems are under intense theoretical and
  experimental  investigation.   Topical  ``fluctuation  relations''  describe
  symmetries  of  the statistical  properties  of  certain  observables, in  a
  variety of  models and  phenomena. They have  been derived  in deterministic
  and,  later, in  stochastic frameworks.   Other results  first  obtained for
  stochastic  processes,  and  later  considered  in  deterministic  dynamics,
  describe the temporal evolution of  fluctuations. The field has grown beyond
  expectation: research  works and different  perspectives are proposed  at an
  ever faster  pace.  Indeed, understanding fluctuations is  important for the
  emerging theory  of nonequilibrium phenomena,  as well as  for applications,
  such as  those of nanotechnological and biophysical  interest.  However, the
  links among the different approaches and the limitations of these approaches
  are not  fully understood.   We focus on  these issues, providing:  {\bf a)}
  analysis  of the  theoretical models;  {\bf b)}  discussion of  the rigorous
  mathematical  results; {\bf  c)} identification  of the  physical mechanisms
  underlying the validity of the  theoretical predictions, for a wide range of
  phenomena.
\end{abstract}
\submitto{\NL}
\maketitle
\tableofcontents





\section{Introduction}
\label{sec:intro}

The study  of fluctuations in  statistical mechanics dates back  to Einstein's
1905  seminal work  on the  Brownian motion  \cite{AE05}, in  which  the first
fluctuation-dissipation relation was given, and to Einstein's 1910 paper which
turned Boltzmann's entropy formula in  one expression for the probability of a
fluctuation out of  an equilibrium state \cite{AE10}. Of  the many authors who
continued Einstein's work, we can recall  but a few. In 1927, Ornstein derived
the fluctuation-dissipation relation for the random force acting on a Brownian
particle \cite{LSO27}.  In  1928, Nyquist obtained a formula  for the spectral
densities and correlation functions of  the thermal noise in linear electrical
circuits, in terms of their impedance \cite{HN28}, which applies to mechanical
systems as well.   In 1931, Onsager obtained the  complementary result to that
of Nyquist: the calculation of the transport coefficients from the observation
of the  thermal fluctuations \cite{LO31a,LO31b}.   The fluctuation-dissipation
theorem and the theory of transport coefficients received great impulse in the
1950's, thanks to the works of authors such as Callen, Welton, R.\ F.\ Greene,
\cite{CWG51,CWG52}, and M.\ S.\  Green and Kubo \cite{MSG51,MSG52,MSG54,RK57}. 
In 1953, Onsager and Machlup  provided a natural generalization to fluctuation
paths  of  Einstein's formula  for  the  probability  of a  fluctuation  value
\cite{OM53a,OM53b}. In  1967, Alder and Wainwright discovered  long time tails
in the velocity autocorrelation  functions, which implied the non-existence of
the  self-diffusion  coefficient, in  two  dimensions \cite{AW67}.   Anomalous
divergent behaviour of the transport coefficients was studied also by Kadanoff
and Swift, for  systems near a critical point  \cite{KS68}.  Closely connected
with  the long  time tails  is the  phenomenon of  long range  correlations in
nonequilibrium steady states, which was pointed out and studied by a number of
authors, including  Cohen, Dorfman, Kirkpatrick,  Oppenheim, Procaccia, Ronis,
Spohn  \cite{LRCa,LRCb,Spohnbook}.  Nonequilibrium  fluctuation  theorems have
been  obtained also by  H\"anggi and  Thomas \cite{HT75,HT78}.   The transient
time correlation  function formalism, which  yields an exact  relation between
nonlinear   steady  state   response   and  transient   fluctuations  in   the
thermodynamic  fluxes,  has been  developed  by  Visscher, Dufty,  Lindenfeld,
Cohen,   Evans   and    Morriss   \cite{Viss,DuLi,Coh,EMTTCF}.    Under   some
differentiability  conditions,   Boffetta  et   al.\  and  Falcioni   et  al.\ 
\cite{FIVa,FIVb} obtained  a fluctuation  response relation, which  applies to
states that  can be very  far from equilibrium.  Independently,  Ruelle proved
that  those conditions  are met  by  axiom A  systems, and  obtained the  same
fluctuation response relation \cite{SRBdiff}.

This  necessarily  brief and  incomplete  account  shows  that the  object  of
research has  gradually shifted from  equilibrium to nonequilibrium  problems. 
But  while the  equilibrium theory  can be  considered quite  satisfactory and
complete, the same cannot be said of the nonequilibrium theory, which concerns
a much wider range of phenomena.
 
The 1993 paper by Evans, Cohen  and Morriss \cite{ECM}, on the fluctuations of
the entropy  production rate  of a deterministic  particle system,  modeling a
shearing fluid, provided a unifying  framework for a variety of nonequilibrium
phenomena, under  a mathematical  expression nowadays called  {\em Fluctuation
  Relation} (FR). Then, fluctuation relations for transient states were proved
by  Evans  and  Searles  in 1994  \cite{earlierpapersA,earlierpapersB},  while
Gallavotti  and  Cohen  obtained  steady  state relations  for  systems  whose
dynamics can  be considered to be  Anosov, in 1995 \cite{GCa,GCb}.   The FR is
one example of  the few exact, general results  on nonequilibrium systems, and
extends the  Green-Kubo and Onsager  relations to far from  equilibrium states
\cite{GG96,GR97,ESR}.

The  subject  of the  present  review is  the  FR  for deterministic  particle
systems, with  an eye on open  problems, and on the  interplay of mathematical
and physical investigations. The  connection with FRs for stochastic processes
is also  discussed. Section~\ref{sec:history} summarizes  the history of  FRs. 
Section~\ref{sec:models} illustrates  a class of  deterministic, time reversal
invariant models of nonequilibrium systems,  relevant in the study of FRs, and
reports some new results.  Sections \ref{sec:GC} and \ref{sec:ESR} illustrate,
respectively,  the   mathematical  theory,  developed  for   the  phase  space
contraction rate, and  the physical mechanisms underlying the  validity of FRs
for quantities such  as the energy dissipation rate.   Section \ref{sec:JE} is
devoted to the Jarzynski and Crooks relations and to their connection with the
FRs.  Sections \ref{sec:vzc} and \ref{sec:JL} illustrate, respectively, {some}
stochastic versions of the FRs,  including the Van Zon-Cohen relation, and the
theory developed  by Jona-Lasinio and  collaborators.  Section \ref{sec:tests}
describes  some  numerical and  experimental  tests  of  the FRs.   Concluding
remarks are made in Section~\ref{sec:conclusions}.

\subsection{Prologue}
Why focus on deterministic rather than stochastic FRs? The stochastic approach
seems to produce  easily the same results that  the dynamical approach obtains
with much effort.   Kurchan says that this is the  case because the stochastic
description, commonly  assumed to be a reduced  (mesoscopic) representation of
the  ``chaotic'' microscopic  dynamics,  is free  from  the intricate  fractal
structures  of deterministic  dynamics \cite{Kurchanlast}.   Then, considering
that the  mathematical approach to deterministic FRs  makes assumptions which,
in general,  cannot be directly verified \cite{Kurchanlast},  one may conclude
that the stochastic approach is to  be preferred to the deterministic one.  In
reality,  there are various  reasons to  consider deterministic  systems.  For
instance, fundamental  issues, like irreversibility, can  hardly be understood
within the  framework of  the intrinsically irreversible  stochastic processes
\cite{Kurchanlast}.   Also,  stochastic   descriptions  assume  that  averages
characterize  single  systems.  This  is  justified only  if  the  microscopic
dynamics  are   sufficiently  ``chaotic''   that  the  average   behaviour  is
established  within   mesoscopic  time  scales  \cite{JB02},   as  happens  in
Thermodynamics, thanks to  the interactions among the particles,  and to their
very  large  number.   However,  in  certain circumstances  particles  do  not
interact  or  interact  more  with  their environment  than  with  each  other
\cite{BuniLP,JR06};  the number  of particles  may be  small;  strong external
drivings  may  produce  ordered  phases;   etc.   In  such  cases,  the  local
thermodynamic  equilibrium   is  violated   and  average  behaviours   do  not
characterize single  systems, they only  characterize ensembles.  Furthermore,
the identification  of physical observables  in stochastic processes  is often
affected by  ambiguities. As far  as the microscopic-mesoscopic  connection is
concerned only  a few models,  like the Lorentz gas  \cite{MH,CELS,LNRM}, have
been mapped  into Markov  processes \cite{BSC}, and  the mapping  concerns the
phase space, not the real space.

Adding that certain results obtained for stochastic processes were not obvious
in terms of  reversible equations of motion (cf.\  Section \ref{sec:JL} on temporal 
asymmetries), while results such as the FRs were not obvious in
the  stochastic  description,  we  conclude  that the  deterministic  and  the
stochastic approaches are both necessary  to provide a unifying framework, for
the field of  nonequilibrium physics, and its applications.   In this review,
  we mainly  focus on deterministic FRs,  but we also  discuss their relations
  with the stochastic ones.

We  illustrate  two classes  of  FRs, transient  and  steady  state FRs.   The
transient  FRs concern  the time  dependent response  to external  drivings of
ensembles of systems, or ensembles of experiments, and hold under very general
conditions  (time  reversibility suffices  for  those  obtained  by Evans  and
Searles;  Hamiltonian dynamics  suffices for  the one  derived by  Jarzynski). 
These   relations   have   interesting    applications   in   the   study   of
nanotechnological  devices and  of biological  systems  \cite{bustamante}, and
hold  for  arbitrarily  short  times.   The  steady  state  FRs  have  similar
applications, but are valid only  asymptotically in time, are harder to derive
in deterministic systems, and have been rigorously obtained only for the phase
space   contraction   rate   of   uniformly   hyperbolic   dynamical   systems
\cite{GCa,GCb}.   Nevertheless,  studying the  mechanisms  which underlie  the
validity of the  steady state FRs for physically  interesting observables, one
understands why they hold so  much more generally \cite{ESR2}.  In particular,
extending the  ensemble derivations of  the transient relations,  one realizes
that time  reversibility and  the decay of  the autocorrelation of  the energy
dissipation imply  the validity  of a  wide class of  steady state  FRs.  Some
decay  of correlations  is  always needed  to  reach a  steady  state, and  to
identify the statistics generated by the  evolution of a single system in real
space, with that of an ensemble of  systems in phase space.  It turns out that
the  form of  mixing  required by  the  steady state  FR's  is minimal.   This
approach,  which justifies  also the  physical  time scales  within which  the
steady state FRs can be verified, is similar to the stochastic approach, as it
deals  with the  time evolution  of  probability measures  (determined by  the
Liouville Equation,  instead of  the Master Equation).   Thus, it  leads quite
easily to a number of results and relations, including the FRs.

Like  the deterministic  and  stochastic descriptions  are complementary,  the
mathematical and physical approaches (summarized in Sections \ref{sec:GC}  and
\ref{sec:ESR}) contribute differently to  our understanding  of nonequilibrium
phenomena, and benefit  from each other's investigations, even  if they mostly
proceed  along  distinct,  parallel  paths. For  instance,  {the  mathematical
  approach} is  concerned with the identification of  dynamical systems  which
allow a rigorous  derivation of some kind of FR, for  one phase function. This
approach  may appear  to  be  physically irrelevant,  because  it may  proceed
independently  of  the  nature of  the  dynamical  systems  and of  the  phase
functions under  investigation. Indeed, the Anosov systems,  whose phase space
contraction  rate  obeys one  FR  \cite{GCa,GCb}, do  not  look  {\em per  se}
physically  revealing; they  may  even be  considered  misleading, since  they
conceal  the true  reasons for  a real  object to  obey one  FR.  Nonetheless,
intriguing physics  questions have  been raised by  the mathematics,  like the
(still  open) question  of which  observables  and which  systems of  physical
interest    verify     the    modified    FR     --Eq.(\ref{axiomCFR})--    of
Refs.\cite{BGG,axiomC}, see e.g.\ \cite{GRS,stephennew}.

On the other hand, {the physical approach} is concerned with understanding the
mechanisms  for which  a {\em  particular} observable,  of one  {\em physical}
system, does  obey a  {\em given} FR.  Thus, {derivations  of the FRs  such as
  those  of  Refs.\cite{review,ESR2}, 
which  are  meant  to  provide  this understanding,  may  look  mathematically
uninteresting,  because they rely  on {\em  physical} assumptions,  which look
impossible to prove.   {These assumptions amount to a  sufficiently fast decay
  of certain  correlation functions, which  makes perfect physical  sense, but
  cannot  be  mathematically  established.  Nevertheless,  similarly  to  the
  arguments of \cite{ECM}, they  introduce an intriguing mathematical problem:
  to construct one dynamical system with  one phase function, for which such a
  decay of correlations can be rigorously assessed.}

The  mathematics and  the physics  still stand on very  different grounds. For
instance, the  mathematically trivial transient relations,  like the transient
$\zW$-FR,  the  Jarzynski  and  the  Crooks  relations,  are  physically  very
interesting: they constitute a challenge for experimentalists, and  carry information on
the physical relevance of current models of nonequilibrium physics. Also, they
are  useful  in  the  study  of  nanoscale biological  systems,  in  which  no
sufficiently general guiding principle has been so far firmly established.

Khinchin's  viewpoint on  the  mathematical ergodic  theory  and the  physical
ergodic  hypothesis \cite{Khinchin}, provides  a notable  analogy for  how the
distinct approaches  to the  FRs may prolifically  interact. Sections 4  and 5
elaborate further on these issues.

\section{Concise History of the Fluctuation Relation}
\label{sec:history}

In 1993, Evans, Cohen and Morriss published a seminal paper \cite{ECM}, on the
fluctuations of the dissipated power, or the entropy production rate $\zs$, in
macroscopic systems  close to  equilibrium. In the  model of  \cite{ECM}, this
observable, later  obtained from the  more general Dissipation  Function $\zW$
\cite{review},  defined  in  Section  \ref{sec:ESR}, equals  the  phase  space
contraction rate  $\zL$ \cite{EM},  defined in Section  \ref{sec:models}.  The
authors of \cite{ECM} proposed and tested the following relation:
\begin{equation}
\frac{P_\zt(A)}{P_\zt(-A)} = e^{\zt A}
\label{firstFR}
\end{equation}
where $A$ and  $-A$ are averages of the dissipated power,  divided by $k_B T$,
on evolution  segments of  duration $\zt$, and  $P_\zt$ is their  steady state
probability.    In    analogy    with    the   periodic    orbit    expansions
\cite{POEinECM2a,POEinECM2b},   Eq.(\ref{firstFR})  was   obtained   from  the
``Lyapunov weights'' in the long $\zt$ limit.  Remarkably, Eq.~\eref{firstFR},
does not contain any adjustable parameter.

In  1994, Evans  and Searles  obtained  the firsts  of a  series of  relations
similar to Eq.\eref{firstFR}, which  we call transient $\zW$-FRs, because they
concern      $\zW$     \cite{review,earlierpapersA,earlierpapersB,generalized,
  SE2000,StephenPRE,stephennew},  for  ensembles of  systems  which evolve  in
time.   The  only requirement  for  the transient  $\zW$-FRs  to  hold is  the
reversibility  of  the  microscopic   dynamics.   Because  they  describe  the
fluctuations  of  $\zW$,  these   relations  can  be  experimentally  verified
\cite{WSMSE}.  Evans  and Searles  argued that, in  the long $\zt$  limit, the
transient $\zW$-FRs  become the steady  state $\zW$-FRs, as indicated  by many
tests,                                                                    e.g.\ 
Refs.~\cite{SE2000,romans,GZG,AustJChem,DK,exptssa,exptssb,stephennew,LLP1,TG,BGG}.

In 1995,  Gallavotti and  Cohen provided a  mathematical justification  of the
Lyapunov  weights  of   Ref.\cite{ECM},  introducing  the  Chaotic  Hypothesis
\cite{GCa,GCb,GG-MPEJ,GGrevisited}:

\vskip  5pt \noi  {\bf Chaotic  Hypothesis:} {\it  A  reversible many-particle
  system in a  stationary state can be regarded as  a transitive Anosov system
  for the purpose of computing its macroscopic properties.}

\vskip  5pt \noi  The result  was a  genuine steady  state FR,  which  we call
$\zL$-FR, as it concerns the  fluctuations of the phase space contraction rate
$\zL$.  This quantity  is proportional  to the  energy dissipation  rate  of a
subclass of  Gaussian isoenergetic particle systems, which  includes the model
of \cite{ECM}.

A strong  assumption as  the Chaotic Hypothesis  raises the question  of which
systems of practical  interest are ``Anosov-like'', since almost  none of them
is  actually Anosov.   The answer  of  Ref.\cite{GCa,GCb} is  that the  Anosov
property,  in  analogy with  the  Ergodic  property,  holds ``in  practice''.  
Difficulties with  the physical interpretation of the  $\zL$-FR emerge because
$\zL$, in general, does not have an obvious physical meaning, and because it is
problematic, when not  impossible, to verify the $\zL$-FR close to equilibrium,
even  in   numerical  simulations  where  $\zL$  is   an  accessible  quantity
\cite{SE2000,romans,GZG,DK,ESR}.

In  1996, Gallavotti  showed that  the FR  constitutes an  extension  to (even
strongly)  nonequilibrium  systems of  the  Green-Kubo  and Onsager  relations
\cite{GG96}.

The  $\zL$-FR applies  to dissipative  systems, {\it  i.e.}  to  systems whose
phase space volumes  on average contract. In \cite{EPRB},  Eckmann, Pillet and
Rey-Bellet studied  the steady  state of an  anharmonic chain coupled  to {\em
  infinite}  thermal baths, so  that the  overall system  is non  dissipative. 
They showed that the relevant  rate of entropy production is strictly positive
and obtained heuristically a suitable FR, which was later rigorously proven by
Rey-Bellet and Thomas \cite{RBT}.

Because fluctuations  are not directly observable in  macroscopic systems, but
can be observed in small systems  or small parts of macroscopic systems, a few
attempts   have   been  made   to   derive  a   local   version   of  the   FR
\cite{GRS,TRVopen,AES,CM00}. This issue deserves further investigation.

The first stochastic FR motivated by Ref.\cite{ECM} was obtained by Kurchan in
1998 \cite{Kurchan}.  The  stochastic FRs of Lebowitz and  Spohn \cite{LS}, of
Evans and  Searles \cite{stochasticES}, and of Maes  \cite{CM99} followed. The
stochastic results  of Van Zon  and Cohen, \cite{VZCa,VZCb},  are particularly
interesting for the theory of deterministic systems. The works by Bodineau and
Derrida \cite{BD}, and by Bertini, De Sole, Gabrielli, Jona-Lasinio and Landim
\cite{BDSJLcurrent}  also lead  to stochastic  FRs. Other  generalizations and
extensions of the \LFR and \WFRs  have been produced by different authors, see
e.g.                                                                            
Refs.~\cite{generalized,stochasticES,VZCa,VZCb,GAFLU,RS,GRS,QKurchan,MT,DRM,Mukamel,JW}. 
It is impossible to mention all of them here. The reader is therefore referred
to the cited literature for more information.

The Jarzynski  equality is  a transient relation,  which connects  free energy
differences  between  two  equilibrium  states  to  non-equilibrium  processes
\cite{CJ}. It was  obtained independently of the FRs in  1997.  In 2000 Crooks
derived an equality that combines  the transient FR and the Jarzynski equality
in just  one formula \cite{GK}. Both  the Jarzynski and  the Crooks equalities
concern  evolving  ensembles  of  nonequilibrium states,  rather  than  single
nonequilibrium  stationary  states.  Hatano  and  Sasa, in  2001,  produced  a
relation of  similar kind  \cite{HS01}, developing the  works of  Paniconi and
Oono \cite{PO98}.

The picture would be completed by a  review of the quantum versions of the FR,
but we cannot elaborate also on that.  On the other hand, as is often the case
for  the  objects  of  statistical  mechanics,  quantum  mechanics  introduces
technical difficulties which must  be treated with appropriate techniques, but
do not modify  the conceptual framework.  Therefore, the  interested reader is
referred       to      the       existing       literature,      such       as
\cite{QKurchan,MT,DRM,Mukamel,JW}.

\section{Dynamical models and equivalence of ensembles}
\label{sec:models}

Let a  system constituted  by $N$ classical  particles, in $d$  dimensions, be
described by:
\begin{equation}
\dot{x} 
= G (x) ~; \quad  x = ({\bf q},{\bf p})  \in \mathcal{M} \subset \zR^{2dN} ,
\label{xdot}
\end{equation}
where $\mathcal{M}$  is the phase space,  and $G$ is determined  by the forces
acting on the  system and by the particles  interactions. A dynamical quantity
of  interest, in the  following, is  the phase  space contraction  rate $\zL$,
defined by
\begin{equation}\label{pscrG}
 \zL=-\mbox{div}\, G ~.
\end{equation}
If the  dynamics are discrete,  $x_{n+1}=F(x_n)$, the phase  space contraction
per unit time is given by
{
\begin{equation}
 \zL=-\log J ~, \quad \mbox{with~~} J
=   \left| \frac{\partial  F}{\partial x}\right|
\end{equation}
the Jacobian determinant of $F$.}  For  continuous time, denote by $S^t x$, $t
\in  \zR$, the  solution of  Eq.(\ref{xdot}) with  initial condition  $x$.  An
observable quantity $\bar\mathcal{O}$ is the  time average of a phase function
$\mathcal{O} : \mathcal{M} \rightarrow \zR$
\begin{equation}
\bar{\mathcal{O}}(x) = \lim_{T\rightarrow\infty} \frac{1}{T} \int_0^T \mathcal{O}(S^t x) d t  ~,
\qquad x \in \mathcal{M}
\label{Taverage}
\end{equation}
Computing  such  a  limit  is  exceedingly complicated,  in  general,  but  in
equilibrium the  problem is commonly  solved by the Ergodic  Hypothesis, which
states that
\begin{equation}
\bar{\mathcal{O}}(x) = \frac{1}{\mu (\mathcal{M})} \int_\mathcal{M} \mathcal{O}(y)  \, d \mu (y) 
\equiv \langle \mathcal{O} \rangle_{\mu}                         
\label{EH}
\end{equation}
for a  suitable measure $\mu$, and  for $\mu$-almost all $x  \in \mathcal{M}$. 
Similar relations hold for discrete time evolutions.

Only  a  few systems  of  physical  interest  verify the  strict  mathematical
statements of the  ergodic theory, and there is no hope  that a many particles
system  will  ever  explore  its  phase  space  as  densely  as  suggested  by
Eq.(\ref{EH}).  Nevertheless the Ergodic Hypothesis is successfully applied in
a very  wide range of situations,  because the variables  of physical interest
are but a few, and tend to constants in the large $N$ limit (cf.\ chapter I of
\cite{GGspringer}  and  Ref.\cite{Khinchin}).   This  means that  the  set  of
observables  of interest  is  too small  to  probe true  ergodicity, and  that
different, necessarily partial models of  the same system may be equivalent in
describing its  limited set of physically  interesting properties.  Therefore,
for an  isolated system whose energy $H$  remains within a thin  shell $[E, E+
\Delta  E]$,  it is  justified  to  {\em postulate}  that  $\mu$  is the  {\em
  microcanonical ensemble}; for a closed system in contact with a heat bath at
given temperature,  the {\em canonical ensemble}  is postulated; and  so on. A
posteriori one checks  whether the assumption is valid or  not, and finds that
these classical ensembles are appropriate in very many situations: they can be
used in practice,  for calulations of physically relevant  quantities.  In the
thermodynamic limit ($N$ becomes large at constant density and energy density)
the different ensembles  become equivalent, in the sense  that the averages of
local observables tend to the same values.

\subsection{The models}
Nonequilibrium  systems   in  steady  states  appear  harder   to  treat  than
equilibrium  phenomena,  thus  one  needs  simple models,  to  assess  various
hypothesis.  From  this stand point, Nonequilibrium  Molecular Dynamics (NEMD)
is one  large reservoir  of interesting models,  which have  been successfully
adopted in  the study  of the  rheology of fluids,  polymers in  porous media,
defects in  crystals, friction  between surfaces, atomic  clusters, biological
macromolecules, among a host of other phenomena \cite{AT87,EM,WH91}.  They are
not reliable if  quantum mechanical effects are important,  if the interatomic
forces are too complicated or insufficiently known, if the number of particles
needs to be too large, or the simulations have to be too long; but NEMD models
are otherwise quite successful in  computing transport coefficients, and are a
valid alternative  to a  number of experiments.  In this paper,  the following
models are used:
\begin{equation}
\begin{array}{lll}
\dot{\bf q}_i&=&{\bf p}_i/m+{\mathcal C}_i\cdot{\bf F}^{ext} ~,\\
\dot{\bf p}_i&=&{\bf F}^{int}_i+{\mathcal D}_i\cdot{\bf F}^{ext}-\za_{th}{\bf
  p}_i ~,\\
\end{array}
\label{geneqnsmotion}
\end{equation}
where ${\bf F}^{ext}$  is the external driving, coupled to  the system via the
constants     ${\mathcal     C}_i$     and     ${\mathcal     D}_i$,     ${\bf
  F}^{int}=-\nabla\Phi^{int}$ is  the conservative  force due to  the internal
interactions among the particles, with interaction potential $\Phi^{int}$, and
$i=1,...,N$.   The   term  $\za_{th}{\bf  p}_i$  is   deterministic  and  time
reversible, and is needed to add or remove energy from the system, in order to
reach  a  steady state  \cite{EM}.  It  is not  a  physical  force;  {it is  a
  ``synthetic''  thermostat  that   substitutes  the  very  many,  practically
  impossible to treat, degrees of freedom of a real thermostat.}

For quantities not affected by how energy is removed from the system, the form
of $\za_{th}$ is irrelevant, because susceptibilities of thermal processes are
similar  to susceptibilities  of mechanical  processes  \cite{EM}. {Therefore,
  driving boundaries may be efficiently replaced by fictitious external forces
  and constraints,  for the purpose  of computing transport  coefficients, and
  {\em  ad  hoc}  models  may   be  devised  as  {\em  equivalent}  mechanical
  representations of  both mechanical  and thermal transport  processes.}  The
theory illustrated in  Refs.\cite{EM,WH91,SEC98} guarantees the correctness of
the results obtained via Eqs.~\eref{geneqnsmotion}.

The models which have been mostly used in the study of the FR are derived from
Gauss' principle of least constraint \cite{gauss,lanczos}:

\vskip 5pt
\noindent
{\bf Gauss Principle (1829): }{\it 
Consider $N$ point particles of mass $m_i$, subjected to
frictionless bilateral constraints ${\bf \phi}_i^{(c)}$ and to external
forces ${\bf F}_i$. Among all motions allowed by the constraints, 
the {\bf natural} one minimizes the ``curvature''
$$
C := \sum_{i=1}^N
m_i \left( \ddot{\bf q}_i - \frac{{\bf F}_i}{m_i} \right)^2
= \sum_{i=1}^N \frac{1}{m_i} \left( {\bf \phi}_i^{(c)} \right)^2 ~.
$$
}
\vskip 5pt

\noindent
The  resulting  equations  of   motion  are  Hamiltonian  only  for  holonomic
constraints.  The isokinetic ($IK$) constraint, which fixes the kinetic energy
$K = \sum_i  {\bf p}_i^2 / 2m$, and the  isoenergetic ($IE$) constraint, which
fixes the  internal energy $H_0  = K +  \zF^{int}$, are not holonomic.   For a
system in  an external  electric field ${\bf  E}$, with  $\mathcal{C}_i=0$ and
$\mathcal{D}_i {\bf  F}^{ext}=c_i{\bf E}$, the  $IK$ and the  $IE$ constraints
lead to
\begin{eqnarray}
\label{aIK}
&&\za_{th} = \alpha_{IK} (x) \equiv \frac {1}{2K} \left( {\bf J} \cdot  
{\bf E} + \sum_{i=1}^N \frac{\bf p_i}{m}
\cdot {\bf F}_i^{int} \right)  \quad \mbox{preserves } K ~, \\
&&\za_{th} = \alpha_{IE} (x) \equiv \frac {1}{2K} {\bf J} \cdot {\bf E}  
\qquad \qquad \qquad \quad \qquad \mbox{preserves } H_0 ~,
\label{aIE}
\end{eqnarray}
where ${\bf J} = \sum_{i=1}^N c_i  \dot{\bf q}_i$, is the electric current and
$c_i$  the  electric  charge  of  the $i$-th  particle.   Another  example  of
Eqs.~\eref{geneqnsmotion}  is a popular  model for  shear flows  called SLLOD,
given by
\begin{equation}
\begin{array}{l}
 \dot{\bf q}_i = {\bf p}_i /  m + {\bf i} \gamma \, y_i ~, \qquad
 \dot{\bf p}_i =  {\bf F}^{int}_i - {\bf i} \gamma \, p_{yi}
  - \alpha_{th} {\bf p}_i ~,
\end{array} 
\label{SLLODeqs}
\end{equation}
and
\begin{equation}
\hskip -30pt
\za_{th} = \alpha_{IK} = \frac{\sum_{i=1}^N \left( {\bf F}^{int}_i \cdot {\bf p}_i 
- \gamma p_{xi}p_{yi} \right)}{\sum_{i=1}^N {\bf p}_i^2} ~, ~~ \
\za_{th} = \alpha_{IE} = \frac{- \zg \sum_{i=1}^N p_{xi}p_{yi}}{\sum_{i=1}^N {\bf p}_i^2} ~,
\label{IKsllodalpha}
\end{equation}
\noindent
where $\gamma$  is the  shear rate  in the $y$  direction and  {\bf i}  is the
unitary vector in the $x$ direction.

In the  above examples, $\za_{IE}$  is proportional to the  power dissipation,
divided by  the kinetic  temperature, which, in  macroscopic systems  in local
equilibrium,  is the  entropy  production rate.   Because  $\zL =  -\mbox{div}
(\dot{\q},\dot{\p})$ is  in turn proportional to $\za_{IE}$,  it can similarly
be related to the entropy  production rate. However, this interpretation faces
the  difficultly that  any  real  nonequilibrium steady  state  can hardly  be
considered  isoenergetic.    Indeed,  it  is  not  possible   to  control  the
redistribution among the  internal degrees of freedom, of  the energy given to
the system by the external  drivings. Hence, the direct relation between phase
space contraction  and energy dissipation appears accidental  and of difficult
interpretation.

Depending on the  physical property to be described, other  models are used in
the literature; like e.g.\ isobaric, isochoric, isoenthalpic, constant stress,
etc.\ models.   {We mention the popular {\em  Nos\`e-Hoover thermostat} model}
\cite{NHa,NHb,NHc}, defined by:
\begin{equation}
\dot{\bf q}_i = {\bf p}_i/m ~, \quad
\dot{\bf p}_i = {\bf F}^{int}_i - \zeta{\bf p}_i ~, \quad
\dot{\bf\zeta} = \frac{1}{\zt^2} \left(\frac{K({\bf p})}{K_0} - 1 \right) ~,
\label{NHeq}
\end{equation}
where $K_0$  is the  chosen average  of the kinetic  energy $K({\bf  p})$, and
$\zt$  is  a relaxation  time.  In  the  small $\zt$  limit,  Eqs.~\eref{NHeq}
approximate the $IK$  dynamics, but are more realistic  and generate canonical
distributions,  in  equilibrium,  as  appropriate for  macroscopic  isothermal
systems.

\subsection{Equivalence and non-equivalence of nonequilibrium ensembles}
The  NEMD models  have been  criticized for  their non-Hamiltonian  structure. 
{However,  a  Hamiltonian structure}  is  not to  be  expected  in systems  in
nonequilibrium steady states,  when the thermostat degrees of  freedom are not
included \cite{RUPhysToday}.   Indeed, let a complete $N$-particle  model of a
system and its thermostat consist of Hamiltonian equations written as
\begin{equation}
\dot{x} = \left( \begin{array}{c}
\dot{x}_s \\
\dot{x}_r \end{array} \right) = G(x) = \left( \begin{array}{c}
G_s(x_s,x_r) \\
G_r(x_s,x_r) \end{array} \right) ~, \quad \begin{array}{l}
x_s = ({\bf q}_i,{\bf p}_i)_{i=1}^{N_s} \\
x_r = ({\bf q}_i,{\bf p}_i)_{i=N_s+1}^{N} \end{array}
\label{NSNR}
\end{equation}
where the  subscript $s$  refers to the  $N_s$ particles of  the thermostatted
system,  and the  subscript $r$  refers to  the $N_r=N-N_s$  particles  of the
reservoir. If one is solely interested in the dynamics of the system variables
$x_s$, then the  projected dynamics will be dissipative  as the reservoirs, on
average, remove energy from the  driven system. This is schematically shown in
Figure~\ref{fig1}. The projected dynamics is time reversal invariant, although
it does  not preserve the volumes in  its reduced space.\footnote{Differently,
  in  systems  of non-interacting  particles,  the  projected dynamics  remain
  Hamiltonian.}  Moreover, if the time  reversed evolution is allowed in phase
space, it is also allowed in the projected space.

\begin{figure}
\centering
\psfig{figure=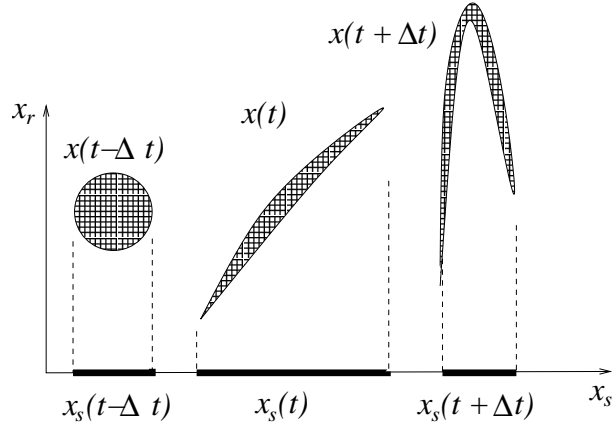,width=8cm,angle=0}
\caption{\small
  Evolution of a phase space  volume in the total, $(x_s,x_r)$, and projected,
  $x_s$,  phase spaces.  In  the  $(x_s,x_r)$ space,  the  dynamics is  volume
  preserving.   The dynamics  projected  on $x_s$  expands  and contracts  the
  volumes.  If the  backward evolution occurs in full phase  space, so it does
  in the projected space.}
\label{fig1}
\end{figure}

Something similar  happens in NEMD models, hence  their non-Hamiltonian nature
is not  a hindrance, by itself. However,  the fact that they  are not obtained
through the ideal projection procedure implies that they must be used {\it cum
  grano salis}: they represent only certain features of nonequilibrium systems
\cite{SEC98,RC98,LRladek,ESA,BTV,GAFLU},             under             certain
conditions.\footnote{For  instance,  large  $N_s$  and  some  form  of  mixing
  produced by  particles interactions is  necessary for the  fictitious forces
  not to dominate the  behaviour of NEMD systems \cite{CR98,JR06}. Furthermore,
  not all kinds of particles  interactions suffice to mimic thermodynamic like
  behaviours \cite{BLnonequiv}.}

To the best of  our knowledge, Refs.\cite{EMequiva,EMequivb} may be considered
the first works on the  equivalence of nonequilibrium ensembles, based on NEMD
models.   For   the  equivalence  of  various   thermostatted  responses,  see
Refs.\cite{LBC,ESA,SEC98,HAHDG,HP04,EM}.   The  papers \cite{HAHDG,HP04}  show
that  the phase  space  dimensionality loss,  due  to dissipation,  is a  bulk
phenomenon even  when the  thermostat acts only  on the  boundaries \cite{AK},
confirming that  boundary thermostats may be replaced,  in some circumstances,
by synthetic  bulk thermostats. References \cite{BTV,MP99} also  deal with the
equivalence of deterministic thermostats.

Nevertheless,  the  equality among  the  entropy  production  rate of  systems
subjected to different  thermostatting mechanisms, as well as  the equality of
this  with the  corresponding  phase  space contraction  rate,  is a  delicate
question.     For    instance,    consider    the   systems    described    by
Eqs.(\ref{geneqnsmotion}) with $\mathcal{C}_i=0$ and constant ${\bf F}^{ext}$,
under  IK and IE  constraints. To  obtain the  equivalence of  their ``entropy
production rates'', one  may proceed as follows \cite{CR98}:  first, note that
the ergodic hypothesis, together with Eq.(\ref{aIK}), yields
\begin{equation}
\hskip -40pt
\overline{\zL}_{_{IK}} = (dN-1) \langle \za_{_{IK}}  \rangle = 
(dN-1) \left[ \left\langle 
\frac{\sum_{i=1}^N \frac{{\bf p}_i}{m} \cdot {\bf F}_i^{int}
}{\sum_{i=1}^N \frac{{\bf p}_i^2}{m}} \right\rangle +
\left\langle \frac{
\sum_{i=1}^N \frac{{\bf p}_i}{m} \mathcal{D}_i \cdot {\bf F}^{ext}
}{
\sum_{i=1}^N \frac{{\bf p}_i^2}{m}
} \right\rangle
\right]
\label{KIK}
\end{equation}
for IK  systems, where the bar  indicates time average and  the brackets phase
space average.   The constraint removes one  degree of freedom,  thus the {\em
  kinetic} temperature $T$ is defined by {
\begin{equation}
(dN-1) k_{_B} T \equiv 2 K = \sum_{i=1}^N \frac{{\bf p}_i^2}{m} = \left\langle 
\sum_{i=1}^N \frac{{\bf p}_i^2}{m} \right\rangle \ ,
\label{TIKss}
\end{equation}
($K$ is constant).
Considering} that the interaction forces do not do any net work,
\begin{equation}
\sum_{i=1}^N \frac{{\bf p}_i}{m} \cdot {\bf F}_i^{int}({\bf q}) = - \frac{d}{d t} \zF^{int}({\bf q}) ~, \quad \mbox{so that }
- \left\langle \frac{d}{d t} \zF^{int} \right\rangle - \frac{d}{d t} \left\langle \zF^{int} \right\rangle = 0
\label{phider}
\end{equation}
and dividing  by the volume $V$  of the system, to  compare dynamical averages
with macroscopic quantities, one obtains:
\begin{equation}
\zs_{IK} \equiv \frac{\langle \zL_{_{IK}} \rangle}{V} = 
\frac{\left\langle \sum_{i=1}^N \frac{{\bf p}_i}{m V}\mathcal{D}_i \cdot 
{\bf F}^{ext} \right\rangle}{k_{_B} T} = \frac{\left\langle \sum_{i=1}^N 
\frac{{\bf p}_i}{m V}\mathcal{D}_i \right\rangle}{k_{_B} T} \cdot 
{\bf F}^{ext} ~.
\label{KIKhat}
\end{equation}
Noting  that   ${\bf  I}  =  \left\langle   \sum_{i=1}^N  \frac{{\bf  p}_i}{m}
  \mathcal{D}_i \right\rangle/V$ is the particle current density, one gets:
\begin{equation}
\zs_{IK} = \frac{{\bf I} \cdot {\bf F}^{ext}}{
k_{_B} T} ~,
\label{KIKhapri}
\end{equation}
where  the  right hand  side  of  Eq.(\ref{KIKhapri})  is {\em  formally}  the
expression   for  the   entropy  production   rate,  $\zs$,   in  Irreversible
Thermodynamics.   In the $IE$  case, there  is no  constraint on  the momenta,
hence the kinetic temperature is defined by:
\begin{equation}
\left\langle \sum_{i=1}^N \frac{{\bf p}_i^2}{m} \right\rangle = d N k_{_B} T ~,
\label{tempdef}
\end{equation}
while Eq.(\ref{aIE}) yields:
\begin{equation}
\hskip -60pt 
\zs_{IE} \equiv \frac{\langle \zL_{_{IE}} \rangle}{V} = 
(dN-1) \left\langle \frac{
\sum_{i=1}^N \frac{{\bf p}_i}{m V} \mathcal{D}_i \cdot 
{\bf F}^{ext}}{
\sum_{i=1}^N \frac{{\bf p}_i^2}{m}} \right\rangle = (dN-1) \left\langle \frac{
\sum_{i=1}^N \frac{{\bf p}_i}{m V} \mathcal{D}_i }{
\sum_{i=1}^N \frac{{\bf p}_i^2}{m}} \right\rangle \cdot 
{\bf F}^{ext}~.
\label{KIEhat}
\end{equation}
For large  $N$, {\it  if one  argues} that the  average of  the last  ratio of
(\ref{KIEhat}) can  be replaced by the  ratio of the  averages --something not
obvious in nonequilibrium systems-- one obtains
\begin{equation}
\zs_{IE} = \frac{{\bf I} \cdot {\bf F}^{ext}}{
k_{_B} T} ~,
\label{KIEhapri}
\end{equation}
up  to  terms of  order  $O(1/N)$.  Therefore,  the  equality  of the  entropy
production rates,  as well as  their equivalence with the  corresponding phase
space contraction rates, for systems with different thermostatting mechanisms,
cannot be taken  for granted, in general, although for properly chosen initial 
conditions, $\langle \zL_{IE} \rangle$ may coincide with $\langle \zL_{IK} \rangle$
in the large $N$ limit. We remark that, without a large number
$N$  of  interacting  particles,  one  could  not  speak  at  all  of  entropy
production.  Indeed, irreversible thermodynamics requires a local equilibrium,
in which the  extensive properties are proportional to  $N$ and depend further
only on the  temperature and on the  number density $n = N/V$.   But for large
$N$, one  could have $\zs_{IE} = \zs_{IK}  + O(1/N)$, in which  case one could
speak of  equivalence of nonequilibrium  ensembles in the  thermodynamic limit
($N,V  \rightarrow  \infty$,  while  density  and energy  density  tend  to  a
constant).    This   idea   has   been   further  developed   by   Ruelle   in
\cite{DRthermolim}.

The  proper  choice  of the  initial  conditions  plays  a  role also  in  the
equivalence   principle   for    hydrodynamics,   formulated   by   Gallavotti
\cite{GAFLU}, which concerns evolution equations like
\begin{equation}
\dot{\bf u} + ({\bf u} \cdot \D) {\bf u} = 
- \frac{1}{\rho} \D p + {\bf g} + \za \Delta {\bf u}
~, \quad \D \cdot {\bf u} = 0 ~,
\label{GNS}
\end{equation}
where, {\bf u} is the fluid  velocity field, $\rho$ the fluid density, $p$ the
pressure,  and  {\bf  g} is  a  constant  force.   If $\za=\nu$  is  constant,
Eq.(\ref{GNS})  is  the Navier-Stokes  (NS)  equation  with  viscosity $\nu$.  
Gallavotti considered the case with
\begin{equation}
\za({\bf u},{\bf \omega},{\bf f}) = 
\frac{\int \left[ \w \cdot {\bf f}   + {\bf \omega} \cdot
({\bf \omega} \cdot \D) {\bf u} \right] ~ d {\bf x}}{
\int \left(\D \times {\bf \omega} \right)^2 d {\bf x}}\,, \label{betau}
\end{equation}
where ${\bf \omega}  = \D \times {\bf u}$ and ${\bf  f}=\D \times{\bf g}$, and
called Eq.(\ref{GNS}) the Gauss-Navier-Stokes (GNS) equation. This equation is
time reversible, and has constant enstrophy $Q = \int \omega^2 d {\bf x}$.  In
periodic  boundary  conditions, expanding  ${\bf  u}$  in  Fourier modes,  and
truncating, yields a dynamical system,  with a certain phase space contraction
rate. Gallavotti then stated the:

\vskip  5pt\noindent   {\bf  Equivalence  Principle.}    {\it  The  stationary
  probability  distributions of  the  NS and  of  the GNS  equations are  {\em
    equivalent} in the limit of  large Reynolds number, provided $Q$ and $\nu$
  are so  related that  the constant  phase space contraction  rate of  the NS
  equation and  the average  of the  fluctuating one of  the GNS  equation are
  equal.}  \vskip 5pt

In analogy with equilibrium  statistical mechanics, this principle is supposed
to hold for local variables, and  the large Reynolds number is invoked for the
fluctuations of $\za$ to be fast  on the observation time scales. Then, if the
average of  $\za$ equals $\nu$, something  that depends on  the initial state,
the behaviour of the NS and the GNS evolutions should be the same.

In Refs.\cite{RS,GRS}, the  Lyapunov spectra of the NS  system, expressed by a
small number (up to 168) of  Fourier modes, were indeed found to coincide with
those of the GNS under different constraints.  This shows that the Equivalence
Principle describes certain dynamical systems related to equations (\ref{GNS},
\ref{betau}),  but  it does  not  answer the  question  of  its relevance  for
turbulence, which  requires simulations  with substantially larger  numbers of
modes.   This  is  still quite  a  demanding  task,  in computational  terms.  
Therefore, for  the cases  of Ref.\cite{GRS} in  which the principle  was best
verified,  we have  increased by  only one  order of  magnitude the  number of
degrees of  freedom, passing from  24 to 440  simulated modes.  The  result is
reported  in Fig.~\ref{Nonequivfig},  where the  spectra corresponding  to the
cases  with  equal  estimated   average  phase  space  contraction  rates  are
represented  by the  thick  lines.  The  spectrum  of the  NS  case has  lower
uncertainty,  since  its  dynamics  fluctuate  less.  Because  there  is  some
uncertainty in the calculation of $\langle  \za \rangle$ in the GNS system, we
show, as  a control test,  two additional spectra  for NS systems,  with quite
smaller and quite bigger $\nu$  than the estimated $\langle \za \rangle$.  The
NS spectra shift, decreasing with  $\zL$ (which is proportional to $\nu$), and
in no case  do they overlap with  the GNS spectrum.  Similarly to  the case of
thermostatted  particle  systems, discussed  above,  this  indicates that  the
Equivalence   Principle  poses   delicate  questions.    In   particular,  its
applicability to models of turbulence deserves further investigation.

\begin{figure}
\begin{center}
\epsfig{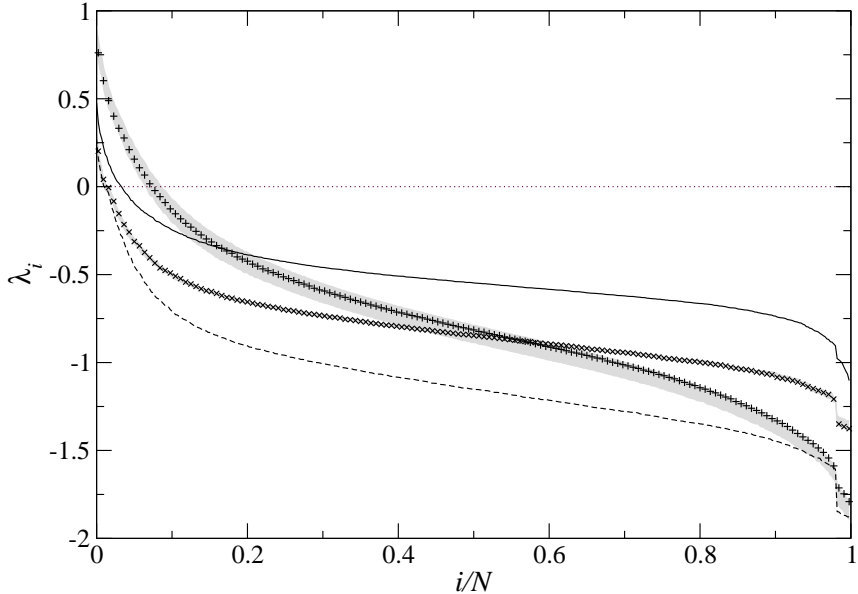}
\caption{
  Lyapunov spectrum  of the  constant energy GNS  system, approximated  by 440
  Fourier modes with $\langle \za \rangle = 0.0105 \pm 0.0006$ (plus symbols),
  and corresponding NS spectrum with  $\nu=0.011$ (cross symbols).  For 
  clarity, only one third of  the symbols  are drawn and
  the  shaded regions around  them correspond  to the  errors, which are estimated  as the
  range between the highest and lowest  computed value, in the last two thirds
  of  the run.  The  upper and  lower thin  lines concern  the NS  system with
  $\nu=0.007$ (solid) and $\nu=0.015$ (dashed), which are well below and above
  $\langle \za \rangle$ (error bars have the same size as for $\nu=0.011$).}
\label{Nonequivfig}
\end{center}
\end{figure}

\section{The mathematical theory}
\label{sec:GC}
The mathematical approach of Gallavotti and Cohen, \cite{GCa,GCb}, is meant to
identify the  context within which  a relation like Eq.(\ref{firstFR})  can be
rigorously derived, i.e.\ to place  on solid mathematical grounds the Lyapunov
weights  used   in  \cite{ECM}.   This  approach  assumes   that  dissipative,
reversible,   transitive   Anosov   diffeomorphisms   are   idealizations   of
nonequilibrium  particle systems,  hence  that the  statistical properties  of
Anosov and  particle systems have  some similarity.  That systems  evolve with
discrete or continuous time, was considered a side issue in \cite{GCa,GCb}, as
apparently confirmed by Gentile's work on Anosov flows \cite{Gentile}.

We now sketch the derivation of the $\zL$-FR of \cite{GCa,GCb}.  Take a smooth
compact manifold $\mathcal{M}$, with a Riemann metric, and a diffeomorphism on
it,  $S  :  \mathcal{M}  \to  \mathcal{M}$,  with  H\"older  continuous  first
derivatives. The dynamical system $(\mathcal{M},S)$ is Anosov if $\mathcal{M}$
is uniformly  hyperbolic for $S$:  i.e.\ there is  a splitting of  the tangent
bundle  $T\mathcal{M} =  V^- \oplus  V^+$, such  that $x  \mapsto  V^\pm_x$ is
H\"older continuous, $TS ~ V^- \subset V^-$, $TS ~ V^+ = V^+$, and
\begin{eqnarray}
&& \| TS^n v \| \le C \theta^{-n} \quad \mbox{ for } v \in V^- \\
&& \| TS^{-n} v \| \le C \theta^{-n} \quad \mbox{ for } v \in V^+
\end{eqnarray}
for all $n \ge 0$ and for given constants $C>0$, $\theta > 1$. The dynamics is
transitive  if  the stable  and  unstable  manifolds  $V^\pm_x$ are  dense  in
$\mathcal{M}$   for   all   $x   \in  \mathcal{M}$.    The   following   holds
\cite{sinaibook}:

\vskip 5pt  \noi {\bf Theorem (Sinai,  1968).  } {\it  Every transitive Anosov
  diffeomorphism has a Markov partition}.  \vskip 5pt

\noi  A Markov  partition is  a subdivision  of $\mathcal{M}$  in  cells whose
interiors are  disjoint from  each other, and  whose boundaries  are invariant
sets constructed  using the  stable and unstable  manifolds.  This  allows the
interior of a cell to be mapped by $S$ in the interior of other cells, and not
across two cells, which would  include a piece of their boundary. Furthermore,
partitions with  arbitrarily small cells can  be constructed. Now,  let $J$ be
the Jacobian  determinant of  $S$, $\zL(x)  = - \log  J(x)$, and  consider the
dimensionless phase  space contraction rate  averaged on a  trajectory segment
$w_{x,\zt}$ of middle point $x \in \mathcal{M}$, and duration $\zt$:
\begin{equation}
e_\zt(x) = \frac{1}{\zt \langle \zL \rangle} \sum_{-\zt/2}^{\zt/2-1} \zL(S^t x) 
\label{p}
\end{equation}
Let $J^u$ be the Jacobian determinant of $S$ restricted to $V^+$.  {\em If the
  system is Anosov}, probability weights of the kind conjectured in \cite{ECM}
can be assigned to the cells of a finite Markov partition, and the probability
that $e_\zt(x)$ falls in  the interval $B_{p,\ze}=(p-\ze,p+\ze)$ coincides, in
the limit  of fine  Markov partitions and  long $\zt$'s,  with the sum  of the
weights  $w_{x,\zt}=\Pi_{k=-\zt/2}^{\zt/2-1}  J^u(S^k  x)^{-1}$ of  the  cells
containing  the   points  $x$  with   $e_\zt(x)  \in  B_{p,\ze}$.    Then,  if
$\pi_{\tau}(B_{p,\ze})$ is the corresponding probability, one can write
\begin{equation}
\pi_\zt(B_{p,\ze}) \approx \frac{1}{M}
 \sum_{x, e_\zt(x)\in B_{p,\ze}} w_{x,\zt}
\label{Piofp}
\end{equation}
where $M$  is a normalization constant.   {\em If the support  of the physical
  measure  is} $\mathcal{M}$,  which is  the case  if the  dissipation  is not
exceedingly high  \cite{ECSB}, time reversibility guarantees  that the support
of $\pi_{\tau}$ is symmetric around $0$, and one can consider the ratio
\begin{equation}
\frac{\pi_\zt(B_{p,\ze})}{\pi_\zt(B_{-p,\ze})} \approx
\frac{\sum_{x,e_\zt(x) \in B_{p,\ze}} w_{x,\zt}}
{\sum_{x,e_\zt(x)\in B_{-p,\ze}} w_{x,\zt}} ~,
\label{pminusp}
\end{equation}
where  each  $x$ in  the  numerator  has a  counterpart  in  the denominator.  
Denoting by  $i$ the  involution which replaces  the initial condition  of one
trajectory with the initial condition of the reversed trajectory,\footnote{For
  instance, $i$  may be the reversal  of momenta, but is  more complicated for
  SLLOD.}  time reversibility yields:
\begin{equation}
\zL(x)=-\zL(ix) ~, \quad w_{ix,\zt} = w_{x,\zt}^{-1} \quad \mbox{and~~~ }
\frac{w_{ix,\zt}}{w_{x,\zt}} = \exp(\zt \langle \zL \rangle p)
\end{equation}
if $e_\zt(x)=p$.  Taking small $\ze$ in $B_{p,\ze}$, the division of each term
in  the numerator  of (\ref{pminusp})  by its  counterpart in  the denominator
approximately equals  $e^{\zt \langle \zL  \rangle p}$, which then  equals the
ratio in (\ref{pminusp}). In the  limit of small $\ze$, infinitely fine Markov
partition and large $\zt$ one obtains:

\vskip  5pt\noindent {\bf  Theorem  (Gallavotti and  Cohen,  1995).} {\em  Let
  $(\mathcal{M},S)$ be dissipative, reversible and chaotic. Then,
\begin{equation}
\frac{\pi_{\tau}(B_{p,\ze})}{\pi_{\tau}(B_{-p,\ze})}
= e^{\zt \langle \zL \rangle p} ~.
\label{largedev}
\end{equation}
with an error in the argument of  the exponential which can be estimated to be
$p,\zt$ independent.  } \vskip 5pt

\noindent
Here, dissipative means  $\langle \zL \rangle > 0$; reversible  means $i S^n =
S^{-n} i$; and  chaotic means that $S$ can be regarded  as a transitive Anosov
system  for  the purpose  of  computing  its  statistical properties  (Chaotic
Hypothesis,  Sec.~\ref{sec:history}).   If  $\zL$  can be  identified  with  a
physical observable, the \LFR is  a statement on the physics of nonequilibrium
systems.

A quite  informative derivation of  the $\zL$-FR is  the one based  on orbital
measures, given by Ruelle in  Ref.\cite{DRdiff}, which we now summarize.  Take
a transitive, reversible Anosov diffeomorphism described above, and a H\"older
continuous  function $A  : \mathcal{M}  \to \zR$;  there is  a  unique ergodic
measure  $\mu$ maximizing  $h(\mu)  + \int  A(x)  \mu(dx)$, where  $h$ is  the
Kolmogorov-Sinai entropy and  $\mu$ is called the Gibbs  state for $A$.  Sinai
proved that the  invariant measure which gives the  forward time statistics is
the Gibbs state of $A=-\log J^u$.  Denoting by Fix$S^n = \{ x\in \mathcal{M} :
S^n x  = x  \}$, the  set of  periodic points of  period $n$  and by  $\phi$ a
continuous function, the probability measure $\mu_n$ defined by the averages
\begin{equation}
\mu_n(\phi) = 
\frac{\sum_{\mbox{Fix} S^n} \phi(x) \Pi_{k=0}^{n-1} J^u(S^k x)^{-1}}
{\sum_{\mbox{Fix} S^n} \Pi_{k=0}^{n-1} J^u(S^k x)^{-1}}
\end{equation}
is  invariant  for  $S$  and  tends  weakly  to  $\mu$  for  $n  \to  \infty$:
$\mu_n(\phi)  \to   \mu(\phi)$  for   all  $\phi$.  Moreover,   by  definition
$\mu(e_\tau) = 1$ and, because of time reversibility, one has
\begin{equation}
J(i S x) = J(x)^{-1} ~, \quad e_\zt \circ i \circ S^\zt = -e_\zt ~, 
\quad \mbox{and ~~~} ~~~ J^s(i S x) = J^u(x)^{-1}
\end{equation}
since $i$  exchanges the stable and  unstable directions.  To  prove the \LFR,
observe first that, given $[p,q] \subset \zR$, there are $a, b > 0$ for which
\begin{equation}
\frac{1}{\zt \langle \zL \rangle} \log 
\frac{\mu(x : e_\zt(x) \in [p,q])}{\mu(x : e_\zt(x) \in [-q-a/\zt,-p+a/\zt])}
\le q + \frac{b}{\zt} ~.
\label{1ststep}
\end{equation}
This rather  sophisticated result  was obtained by  Ruelle relying  heavily on
properties of Anosov  diffeomorphisms, hence it should hardly  be generic (see
also \cite{GGCHAOS}).  Indeed, Eq.(\ref{1ststep}) relies on Bowen's shadowing,
topologically  mixing, the  specification  property, a  property  of sums  for
H\"older  continuous  functions,  the   expansiveness  of  the  dynamics,  the
continuity of the splitting of the tangent bundle, and the expression of $\mu$
in  terms   of  periodic   orbits  ({\em  c.f.},   Sections  3.1  to   3.8  of
Ref.~\cite{DRdiff}). Furthermore, Ruelle uses a large deviation result for one
dimensional systems  with short  range interactions, considering  $A =  - \log
J^u$   and    $B   =   -(1/\langle    \zL   \rangle)   \log   J$,    so   that
$(1/\zt)\sum_{k=0}^{\zt-1}  B(S^k x)  =  e_\zt(x)$.  This  result states  that
there is a real analytic and  strictly concave function $\eta$ in the interval
$(-p^*,p^*)$  such  that,  for  every  other  interval  $I$  which  intersects
$(-p^*,p^*)$, the following holds
\begin{equation}
\lim_{\zt \to \infty} \frac{1}{\zt \langle \zL \rangle} \log \mu
\left( \left\{ x : e_\zt(x) \in I \right\} \right) = 
\frac{1}{\langle \zL \rangle} \sup_{u \in I \cap (-p^*,p^*)} \eta(u)
\end{equation}
where the  Gibbs state of $A$, $\mu$,  and time reversibility have  been used. 
Combining this with (\ref{1ststep}), one obtains
\begin{equation}
\lim_{\zt \to \infty} \frac{1}{\zt \langle \zL \rangle} \log
\frac{\mu(\{ x : e_\zt(x) \in (p-\zd,p+\zd)\})}
{\mu(\{ x : e_\zt(x) \in (-p-\zd,-p+\zd)\})} \le p+\zd
\label{prethm0}
\end{equation}
for $\zd>0$ and $|p| < p^*$.  Taking this result, and the one corresponding to
$-p$, the Gallavotti-Cohen fluctuation theorem is finally obtained.

\vskip 5pt \noi  {\bf Theorem (Ruelle, 1999).} {\it Let  $S$ be a $C^{1+\za}$,
  $\za>0$,   Anosov   diffeomorphism  of   the   compact  connected   manifold
  $\mathcal{M}$,  and  let $\mu$  be  the  corresponding  SRB measure.  Assume
  reversibility,  with involution  $i$, and  consider the  dimensionless phase
  space  contraction rate  $e_\zt$ with  respect to  an  $i$-invariant Riemann
  metric on $\mathcal{M}$. Then, there exists $p^* > 0$ such that the \LFR
\begin{equation}
p - \zd \le \lim_{\zt \to \infty} \frac{1}{\zt \langle \zL \rangle} \log
\frac{\mu(\{ x : e_\zt(x) \in (p-\zd,p+\zd)\})}
{\mu(\{ x : e_\zt(x) \in (-p-\zd,-p+\zd)\})} \le p+\zd
\label{prethm}
\end{equation} 
holds if $|p| < p^*$ and $\zd > 0$.}
\vskip 5pt
\noi

\subsection{Consequences of the $\zL$-FR}
Taking  $\zL$ as  the entropy  production rate,  Gallavotti used  the  \LFR to
obtain  Green-Kubo like  and Onsager  like relations,  in the  limit  of small
dissipation \cite{GG96}.  This  way, the \LFR appears as  an extension of such
relations to  nonequilibrium systems. Gallavotti assumes  that the (reversibly
thermostatted, continuous  time) system is  driven by the $s$  fields $F=(F_1,
F_2,...,F_s)$,  that  $\zL$  vanishes  for  $F=0$, that  the  phase  space  is
bounded,\footnote{Generic  reversible  thermostats  (Ref.\cite{GG96}  mentions
  Gaussian  isokinetic, Gaussian  isoenergetic and  Nos\'e-Hoover thermostats)
  seem to contrast with the requirement of vanishing $\zL$ at equilibrium, and
  with the boundedness of the phase space. In fact, difficulties occur in this
  approach, if it  is applied to NEMD models  (cf.\ Ref.\cite{ESR} and Section
  4.3 below). However,  Ref.\cite{GG96} is not to be  interpreted as referring
  to concrete  particle systems;  it refers to  $\zL$ for  hypothetical Anosov
  systems, provided they exist \cite{GGrevisited}. }  and that
\begin{equation}
\zL(x) = \sum_{i=1}^s F_i J_i^0(x) + O(F^2) ~,
\end{equation}
where $J_i^0$ are the currents close to equilibrium, {\it i.e.}, are linear in
$F$.  Then, the  fast decay of the $\zL$-autocorrelation  function, implied by
the Anosov property, leads to
\begin{equation}
\zeta(p) = - \lim_{\zt \to \infty} \frac{1}{\zt} \log \pi_\zt(p) = 
\frac{\langle \zL \rangle^2}{2 C_2} (p-1)^2 + O((p-1)^3 F^3)
\end{equation}
where $C_2 = \int_{-\infty}^\infty \langle \zL(S^t \cdot) \zL(\cdot) \rangle^T
dt$, and $\langle . \rangle^T$ denotes the cumulant. Thus, using the $\zL$-FR,
one  obtains $\langle \zL  \rangle =  C_2/2 +  O(F^3)$.  Arbitrarily  far from
equilibrium,  Gallavotti   defines  the  currents   as  $J_i(x)=\partial_{F_i}
\zL(x)$, and  the transport coefficients  as $L_{ij} =  \partial_{F_j} \langle
J_i  \rangle |_{F=0}$.   The derivatives  with respect  to the  parameters $F$
require a property of differentiability of SRB measures, which has been proven
by Ruelle in Ref.\cite{SRBdiff}.  Assuming  this property, the validity of the
\LFR and using time reversibility, one can write $\partial_{F_j} \langle J_i^0
\rangle|_{F=0} = \partial_{F_j} \langle J_i \rangle|_{F=0}$, and
\begin{equation}
\hskip -40pt
\frac{1}{2} \int_{-\infty}^\infty  \langle  \zL(S^t x) \zL(x)
\rangle^T dt = \langle \zL \rangle = \frac{1}{2} \sum_{i,j=1}^s (\partial_{F_j} 
\langle J_i \rangle + \partial_{F_i} \langle J_j \rangle) |_{F=0} F_i F _j ~,
\end{equation}
in  the limit  of small  $F$.   Then, if  $s=1$, one  recovers the  Green-Kubo
relations  for  the unique  transport  coefficient  $L_{11}$.   To obtain  the
Onsager  symmetry $L_{ij}=L_{ji}$,  Gallavotti extends  the \LFR  in  order to
consider the joint distribution of  $\zL$ and its derivatives. Introducing the
dimensionless current $q$ in a trajectory segment
\begin{equation}
q(x) = 
\frac{1}{F_j \langle \partial_{F_j} \zL \rangle \zt} 
\int_{-\zt/2}^{\zt/2} F_j \partial_{F_j} \zL(S^t x) d t
\end{equation}
and the joint distribution  $\pi_\zt(p,q)$, with corresponding large deviation
functional  $\zeta(p,q)   =  -   \lim_{\zt  \to  \infty}   \frac{1}{\zt}  \log
\pi_\zt(p,q)$, one obtains a relation like the \LFR:
\begin{equation}
\lim_{\zt \to \infty} \frac{1}{\zt \langle \zL \rangle p} \log
\frac{\pi_\zt(p,q)}{\pi_\zt(-p,-q)} = 1 ~.
\label{jointpq}
\end{equation}
This makes  the difference $(\zeta(p,q)  - \zeta(-p,-q))$ independent  of $q$,
and leads to  the desired result, $L_{ij}=L_{ji}$, in the limit  of small $F$. 
This work inspired Refs.\cite{GR97,RC98,LRladek}. 

The above  derivations are valid if  the dynamics is transitive,  i.e.\ if the
dissipation is not  too high. It is  very hard to violate this  condition in a
particle system \cite{ECSB}. However,  this possibility has been considered in
\cite{axiomC},  where a stronger  hypothesis than  the Chaotic  Hypothesis has
been  introduced, under  the assumption  that the  Lyapunov exponents  come in
pairs that  sum to a constant  $c<0$ \cite{ECM1,DMpairinga,DMpairingb}, except
for some pair of vanishing exponents.  In Ref.~\cite{BGG}, Bonetto et al.\ had
conjectured that the \LFR should generalize to the form
\begin{equation}
\log \frac{\pi_\zt(B_{p,\ze})}{\pi_\zt(B_{-p,\ze})} = \zt \langle \zL_r \rangle
\frac{D-M}{D} p \ ,
\label{axiomCFR}
\end{equation}
apart from  small errors.  To  understand the meaning  of $(D-M)/D \le  1$ and
$\zL_r$,  consider  transitive  dynamics,  and neglect  the  trivial  Lyapunov
exponents. Half of the remaining  exponents are positive ($\zl_i^+$), half are
negative ($\zl_i^-$), and can  be arranged in $D$ pairs $\{\zl_i^+,\zl_i^-\}$,
with $c_i =  \zl_i^++\zl_i^-$. According to the authors  of \cite{BGG}, as the
dissipation grows, the dynamics ceases  to be transitive, some of the positive
exponents become negative and  lower dimensional attracting and repelling sets
are  generated.   If  conjugate  pairing  holds,  i.e.   if  $c_i=c$  for  all
$i=1,...,D$, it  could happen that the  volume contraction along  each pair of
directions corresponding to each pair of exponents is proportional to $c$, and
that the dimensionality of the  attracting manifold $\mathcal{M}_r$ is that of
$\mathcal{M}$  minus the  number $M$  of pairs  with two  negative  exponents. 
Then,  $\zL$ equals  $(D-M)/D$ times  the contraction  rate restricted  to the
attractor, $\zL_r$  and, if  the attractor is  invariant with respect  to some
kind of  time reversal operation, the  FR holds for $\zL_r$,  while $\zL$ must
obey Eq.(\ref{axiomCFR}).\footnote{The required time reversal operation is one
  involution  $i$, obeying  $i  S^t =  S^{-t}  i$, that  leaves the  attractor
  invariant.}

Eq.(\ref{axiomCFR}) is hard to  test in particle systems, because fluctuations
become less and less frequent as the dissipation grows. In Ref.\cite{BGG}, the
case with $(D-M)/D = 18/19$ was not distinguishable from 1, given the achieved
resolution, while the case with $(D-M)/D = 17/19$ could not be tested; similar
difficulties   were   met  in   \cite{GZG}.   An   indirect  confirmation   of
Eq.(\ref{axiomCFR}),  based however on  new scaling  assumptions, is  given in
\cite{RM03}, where a  procedure is given to estimate  finite $\zt$ corrections
to the  steady state  $\zL$-FR. Differently, Ref.\cite{stephennew}  finds that
the standard \LFR holds for a simple oscillator model, even in the presence of
pairs of  negative Lyapunov  exponents.  The theory  of Refs.\cite{BGG,axiomC}
has been generalized to hydrodynamic  models, where conjugate pairing does not
hold,  but fluctuations  persist even  with a  substantial excess  of negative
Lyapunov  exponents  \cite{GAFLU,RS,GRS}.   The  factor  $(D-M)/D$  was  there
replaced by
\begin{equation}
c = \frac{\sum^* (\zl_k + \zl_{2K-1-k})}{\sum (\zl_k + \zl_{2K-1-k})} \ ,
\label{pairs}
\end{equation}
where the $2K-2$  nontrivial Lyapunov exponents are given  in decreasing order
($\zl_1 \ge \zl_2 ... \ge  \zl_{2K-2}$), and $\sum^*$ means summation over the
pairs with one  positive exponent, while $\sum$ is the  summation over all the
pairs.  The \LFR  with  slope $c$  defined  by (\ref{pairs})  was verified  in
GNS  systems truncated  to few  tens of  modes  \cite{GRS}. 

\subsection{Local fluctuations}
In  Ref.\cite{GRS}  a  local  version  of  the  $\zL$-FR,  first  proposed  in
\cite{GGlocal}, was also tested.  One  reason for developing local FRs is that
global  fluctuations are  not  observable in  macroscopic  systems. The  local
$\zL$-FR  of   Ref.\cite{GGlocal}  concerns   an  infinite  chain   of  weakly
interacting chaotic maps.  Let $V_0$ be  a finite region of the chain centered
at the origin, $T_0 > 0$ be a time interval, and define
\begin{equation} \langle \zL \rangle \equiv \lim_{V_0,T_0 \rightarrow
\infty} \frac{1}{|V_0| T_0} \sum_{j=0}^{T_0-1} \zL_{V_0} (S^j x) ~,
\quad p = \frac{1}{\langle \zL \rangle |V|} \sum_{j=-T_0/2}^{T_0/2} 
\zL_{V_0} (S^j x) ~,
\label{etaplus}
\end{equation}
where $V  = V_0 \times T_0$,  and $\zL_{V_0}(x)$ is the  contribution to $\zL$
given by $V_0$.  Then, one obtains:
\begin{equation}
\pi_V(p) = e^{\zeta(p)|V| + O(|\partial V|)} ~, \quad
\mbox{with } \frac{\zeta(p) - \zeta(-p)}{p \langle \zL \rangle } = 1 \quad \mbox{and }
|p| < p^* ~,
\label{pilocal}
\end{equation}
where $|\partial  V|$ is  the size  of the boundary  of $V$,  $p^* \ge  1$ and
$\zeta$ is analytic  in $p$. The contribution of  the boundary term $|\partial
V|$ decreases with growing $V$, leading  to the $\zL$-FR in the limit of large
(compared to microscopic scales) volume $V_0$ and long times $T_0$.

The  problem of  local fluctuations,  naturally  leads to  the possibility  of
extending Onsager-Machlup theory to nonequilibrium systems. This has been done
by Gallavotti \cite{GG-OM1}, under  the assumption that the entropy production
rate is proportional to  $\zL$.  The Onsager-Machlup theory \cite{OM53a,OM53b}
concerns the paths of small  fluctuations around equilibrium states, and leads
to  a   derivation  of  the  hydrodynamic  equations   for  the  corresponding
observables, via the maximization of  the probability of the relaxation paths,
in the  large system limit \cite{OM53a,OM53b}.  Gallavotti  also considers the
probability  of  temporal  paths   $t  \mapsto  \varphi(t)$,  for  observables
$\mathcal{O}$ which are  either even or odd with respect  to the time reversal
operation, and have  vanishing mean. The fluctuation $\varphi$,  is assumed to
be smooth and to  vanish for large $|t|$, but no bound is  placed on its size. 
One may then consider the probability that $\mathcal{O}(S^t x)$ stays close to
$\varphi$,  in the  time interval  $[-\zt/2,\zt/2]$, and,  in the  large $\zt$
limit, one may consider the large  deviation function for $\zL$ to take values
close   to  $p$   and  for   $\mathcal{O}$   to  stay   close  to   $\varphi$,
$\zeta(p,\varphi)$ say.

The   result  is   that  the   path   $\varphi(t)$  and   its  time   reversal
$i\varphi(t)=\pm  \varphi(-t)$ (where  $+$  holds  for even  and  $-$ for  odd
$\varphi$)  are  followed  with  equal   probability  if  the  first  path  is
conditioned to an average $\zL$ equal to $p$ and the second path to an average
$\zL$  equal  to $-p$.   Indeed,  for  time  reversal invariant,  dissipative,
transitive Anosov systems, Gallavotti obtains
\begin{equation}
\frac{\zeta(-p,i\varphi) - \zeta(p,\varphi)}{p \langle \zL \rangle} = 1 ~,
\qquad |p| \le p^* ~, \qquad p^* \ge 1 ~,
\label{GG-OMrel}
\end{equation}
which means  that it suffices  to make $\zL$  behave strangely (i.e.\  to take
values different  from $\langle \zL  \rangle$), to see all  observables behave
equally strangely.   The further developments of  Ref.\cite{GG-OM2} excluded a
direct connection of these results with the theory of Ref.\cite{BDSJLcurrent}.

\subsection{Applicability of the Chaotic Hypothesis}
The question  arises of whether any  system of physical  interest verifies the
Chaotic Hypothesis. As  the FR implied by the  Chaotic Hypothesis concerns the
physically  non-obvious $\zL$,  this question  has been  little  investigated. 
Some of  the papers in  which $\zL$ was  considered suggested that  the steady
state  $\zL$-FR  holds for  reversible  dynamical  systems  with one  or  more
positive  Lyapunov exponents \cite{GG96},  but also  for some  systems without
positive exponents \cite{LRB,BR01}.  That $\zL$ should be bounded and that the
\LFR should hold only for $|p| \le  p^*$, for some $p^*>0$, was not thought to
have observable consequences, at first.

Later it was realized  that the \LFR is hard, if not  impossible, to verify in
non-isoenergetic   systems    in   steady   states    close   to   equilibrium
\cite{SE2000,DK,romans}, despite {the ``higher chaos'' of equilibrium} states.
To  explain these  facts, Ref.\cite{ESR}  observes  that the  \LFR implies  an
asymmetry between positive and negative  fluctuations, which is not present in
equilibrium, hence that the \LFR for non-normalized $\Lambda$ may hold only if
its domain  tends to  $\{0\}$ when  the steady state  tends to  an equilibrium
state.   In the  Gaussian isokinetic  case,  however, $\zL$  is the  sum of  a
dissipative term,  $\zW$, and  a conservative interaction  term, which  may be
singular,  (cf.\  Eqs.~\ref{aIK},\ref{IKsllodalpha}).   The  dissipative  term
obeys the FR, while the conservative term does not, but its averages over long
time intervals are small, and become negligible with respect to those of $\zW$
as the  intervals grow  \cite{romans,ESR}.  Thus, in  the long time  limit the
\LFR may hold  as a consequence of  the validity of the \WFR  \hskip -4pt, but
its  convergence  times diverge  as  the  steady  states approach  equilibrium
states.   Moreover, the  convergence  of the  domain  of the  \LFR to  $\{0\}$
implies that  the \LFR eventually  describes only trivial  fluctuations.  This
causes some difficulty in the  derivation of the Green-Kubo relations from the
$\zL$-FR,  which  requires  the  equilibrium  limit.  On  the  one  hand,  the
averaging times have to  be long for the Central Limit and  the \LFR to apply,
{but  not so  long that  $\zW=0$}  is in  the tails  of the  $\zW$-probability
distribution function, which  are not described by the  Central Limit Theorem. 
If the averaging time required by  the \LFR tends to infinity, this compromise
may  not be  possible.   Singularities of  $\zL$,  in turn,  make dubious  the
existence  of the  cumulants  used  in \cite{GG96}  to  derive the  Green-Kubo
relations.   Therefore, the  physical  applications  of the  \LFR  and of  the
Chaotic Hypothesis appear problematic from this point of view.

References  \cite{ESR,DJE03} suggested that,  in IK  systems, $\zL$  is better
suited to describe  heat fluxes than entropy productions,  hence that the \LFR
has  to be  modified like  the heat  FR of  Van Zon  and Cohen  for stochastic
systems \cite{vzc}.  Indeed, for continuous time systems  with singular $\zL$,
terms  of  the form  $[\zF^{int}(S^\zt  x)-\zF^{int}(x)]/\zt$, with  unbounded
interaction potential ${\zF}^{int}$, affect the large deviations  of $\zL$, if
the probability  distribution of $\zF^{int}$  has exponential or  larger tails
\cite{BGGZ}. The  solution of Ref.\cite{BGGZ} consists in  assuming that chaos
due to  uniform hyperbolicity  may play the  same role  as the white  noise in
Ref.\cite{VZCa,VZCb}.  In the Gaussian isokinetic, or Nos\'e-Hoover isothermal
cases, one has
\begin{equation}
\dot{\q}_i = \p_i \qquad \dot{\p}_i = E - \partial_{\q_i} \zF^{int} -\za \p_i
\qquad \zL = \zL^{(0)} - \zb \dot{V}
\end{equation}
where  $V$  is  related  to   $\zF^{int}$,  and  has  an  equilibrium  ($E=0$)
distribution with exponentially decaying tails, while $\zL^{(0)}$ has Gaussian
tails. It is then assumed that the  tails have same properties when $E \ne 0$. 
Then,  the  average of  $\zL$  in  a time  $\zt$  takes  the  form
{
\begin{eqnarray}
  &&\overline{\zL}_{0,\zt}(x)  = \frac{1}{\zt}  \int_0^\zt  \zL(S^t x)  d t  =
  \overline{\zL^{(0)}}_{0,\zt}(x) +
  \frac{\zb}{\zt} \left[ V(S^\zt x) - V(x) \right] ~,  \\
  \hskip  -50pt  \mbox{with~~~  } \qquad  &&\overline{\zL^{(0)}}_{0,\zt}(x)  =
  \frac{1}{\zt} \int_0^\zt \zL^{(0)}(S^t x) d  t \ ,
\end{eqnarray}
and} for large  $\zt$, in some cases, one may  assume that $\zL^{(0)}$, $V_f=V
\circ S^\zt$ and $V_i=V$ are independently distributed. {This ultimately leads
  to \cite{BGGZ}}
\begin{eqnarray}
&&\hskip -50pt
\lim_{\zt \to \infty} \frac{1}{\zt} \log
\int_{-p^* \langle \zL \rangle}^{p^* \langle \zL \rangle}
d \zL^{(0)} \int_0^\infty d V_i \int_0^\infty d V_f
e^{\zt \tilde{\zeta}_0(\zL^{(0)}) -\zb(V_i+V_f)} \zd[\zt(\zL-\zL^{(0)})+
\zb(V_i - V_f)] \\
&& = \lim_{\zt \to \infty} \frac{1}{\zt} \log 
\int_{-p^* \langle \zL \rangle}^{p^* \langle \zL \rangle} d \zL^{(0)}
e^{\zt \tilde{\zeta}_0(\zL^{(0)}) - \zt | \zL - \zL^{(0)} |} \ ,
\end{eqnarray}
where $\tilde{\zeta}_0(\zL^{(0)})$ is the rate function of $\zL^{(0)}$.  Then,
for the rate function of $\zL$ one obtains
\begin{equation}
\hskip -60pt
\tilde{\zeta}(\zL) = 
\max_{\zL^{(0)} \in [-p^* \langle \zL \rangle,p^* \langle \zL \rangle]}
\left[ \tilde{\zeta}_0(\zL^{(0)}) - | \zL - \zL^{(0)} | \right] = \left\{
\begin{array}{lll} \tilde{\zeta}_0(\zL_-) - \zL_- + \zL &  ; \ & \zL < \zL_- \\
\tilde{\zeta}_0(\zL) &  ; \ &  \zL_- \le \zL \le \zL_+ \\
\tilde{\zeta}_0(\zL_+) + \zL_+ - \zL &  ; \ &  \zL > \zL_+
\end{array}
\right.
\end{equation}
where $\tilde{\zeta}_0'(\zL_{\pm}) = \mp 1$.  If the FR holds for $\zL^{(0)}$,
with  $| \zL^{(0)} |  \le p^*  \langle \zL  \rangle$, the  $\zL^{(0)}$-FR, one
obtains
\begin{equation}
\tilde{\zeta}(\zL) - \tilde{\zeta}(-\zL) = 
\left\{ \begin{array}{lll} \zL &   ; \ & 0 \le  \zL  < \langle \zL \rangle \\
\tilde{\zeta}_0(\zL) + \zL  &  ; \  &  \langle \zL \rangle \le  \zL  \le \zL_+ \\
\tilde{\zeta}_0(\zL_+) + \zL_+ &  ; \ &   \zL  > \zL_+ 
\end{array}
\right. \ .
\label{L_0}
\end{equation}
A relation similar  to the heat FR of  Van Zon and Cohen is  thus obtained for
$\zL$.  The  statement that  the $\zL^{(0)}$-FR holds  with $|p| \le  p^*$, if
$\zL^{(0)}$  is  bounded  or  decays  faster  than  exponential  is  justified
adopting
Gentile's approach  for Anosov flows, which  reduces the flow  to a Poincar\'e
map \cite{Gentile}, and assuming the Chaotic Hypothesis for the resulting map.
In particular, the dynamics may  be restricted to a level surface $V=\bar{V}$,
with $\bar{V} < \infty$, so  that the volume contraction rate, $\zL^{(0)}$, is
bounded and the terms $(V_f - V_i)$ vanish.

This  scenario is  supported by  Gilbert's Ref.\cite{TG},  for a  one particle
system.   However, all other particle systems have been  found to  satisfy the
original $\Lambda$-FR, suggesting that the singularities in the potential term
may not  be sufficient for the  validity of the heat  FR of Van  Zon and Cohen
\cite{MMR1}.  For stochastic systems,  the first indication that singularities
may  invalidate  the \LFR  is  found  in \cite{Farago};  Ref.\cite{Maesrecent}
suggests that the Van Zon-Cohen FR may have quite wide applicability, see also
\cite{PRV}, while \cite{germans} shows some counterexample.  The phenomenology
is quite complex, as Visco explains  \cite{visco}, hence it is not possible at
present to draw the limits of validity of the theory of \cite{BGGZ}.

The above  shows that the \LFR  rests on strong assumptions,  which are hardly
met  by  systems of  physical  interest, and  which  have  no simple  physical
interpretation. At  the same  time, the physically  more obvious  steady state
\WFR is quite generally verified, and does not incur in the difficulties which
affect the $\zL$-FR. Thus  the mathematical theory raises intriguing questions
for the  physical theory: does  the \WFR hold  independently of the  $\zL$-FR? 
Which are  the physical  mechanisms underlying the  validity of  the $\zW$-FR,
when    it    holds?    It    would    also    be    interesting    to    test
Eqs.(\ref{jointpq},\ref{GG-OMrel})  in  NEMD  models,  as well  as  in  actual
experiments,         and          it         is         desirable         that
Eqs.(\ref{axiomCFR},\ref{pilocal},\ref{L_0}) be further investigated.

\section{The physical mechanisms}
\label{sec:ESR}

In 1994, Evans and Searles obtained the first of a series of relations similar
to  Eq.(\ref{firstFR}), for the  {\em Dissipation  Function} $\zW$,  which, in
nonequilibrium states  close to  equilibrium can be  identified with  the {\em
  entropy production rate}, $\zs = J V  F^{ext} / k_{_B} T$.  Here, $J$ is the
(intensive) flux due  to the thermodynamic force $F^{ext}$,  $V$ is the volume
and  $T$ the  kinetic temperature  \cite{earlierpapersA,earlierpapersB}.  That
relation,  called   transient  $\zW$-FR,   is  obtained  under   virtually  no
hypothesis, except  for {\em time  reversibility}; it is transient  because it
concerns non-invariant ensembles of systems, instead of the steady state.  The
transient       $\zW$-FR      has      been       verified      experimentally
\cite{CRWSSE,exptssa,exptssb}, and its  conjectured extension to steady states
has been  validated by many tests.   The Evans-Searles approach  to the steady
state  $\zW$-FR is  based on  the belief  that the  complete knowledge  of the
invariant  measure  implied  by  the  Chaotic  Hypothesis  is  not  needed  to
understand a few properties of the steady state.  Like thermodynamic relations
are widely applicable because do not  depend on the details of the microscopic
dynamics,  the  observed  wide  applicability  of the  steady  state  $\zW$-FR
suggests, indeed,  that it  cannot depend on  subtle dynamical  features, like
approximate  hyperbolicity.  It  is  therefore  necessary  to  understand  the
mechanisms underlying the validity of  the steady state $\zW$-FR in systems of
physical interest.

Following Ref.\cite{ESR2}, let ${\cal M}$ be  the phase space of the system at
hand, and $S^\tau: {\cal M} \rightarrow {\cal M}$, a reversible evolution with
time reversal map $i$. Take a  probability measure $d \mu(\zG) = f(\zG) d \zG$
on ${\cal M}$, and let the observable $\mathcal{O} : {\cal M} \rightarrow \zR$
be  odd with  respect to  time reversal  {\it i.e.},  \ $\mathcal{O}(i  \zG) =
-\mathcal{O}(\zG)$. Denote its time averages by
\begin{equation}
\Ft(\zG) \equiv \frac{1}{\tau} \mathcal{O}_{t_0,t_0+\zt}(\zG) =
 \frac{1}{\tau} 
\int_{t_0}^{t_0+\tau} \mathcal{O}(S^{s} \zG) d s ~.
\label{phitau}
\end{equation}
For a  density $f$ even with  respect to time reversal,  {\it i.e.} satisfying
$f(i \zG)=f(\zG)$, define the {\em Dissipation Function} as
\begin{equation} \label{omegat}
\hskip -65pt
\zW(\zG)  = - \left. \frac{d}{d \zG} \log f
\right|_\zG   \cdot  \dot{\zG}   +  \zL(\zG)  ~, \quad \mbox{so that} ~~~
\Wt(\zG) = \frac{1}{\zt}
\left[ \ln \frac{f(S^{t_0}\zG)}{f(S^{t_0+\zt} \zG)} +
\zL_{t_0,t_0+\tau} \right]
\end{equation}
{Note   that,   for  a   compact   phase   space},   the  uniform   density
$f(\zG)=1/|{\cal M}|$ implies $\zW=\zL$.  However, $\zW$ equals the dissipated
power, divided by the kinetic temperature, in bulk thermostatted systems, like
those of Eqs.\eref{geneqnsmotion}, only  if $f$ is the equilibrium probability
density for  the given system  \cite{ESR2}, and only in  special circumstances
does this  imply $f(\zG)=1/|{\cal M}|$.   That the logarithmic term  exists in
(\ref{omegat})  has been  called  {\em ergodic  consistency} \cite{review},  a
condition met if $f>0$ in all regions visited by the trajectories $S^t \zG$.

For $\zd > 0$,  let $A^+_\zd=(A-\zd,A+\zd)$ and $A^-_\zd=(-A-\zd,-A+\zd)$, and
let  $E(\mathcal{O}  \in  (a,b))$  be  the  set  of  points  $\zG$  such  that
$\mathcal{O}(\zG) \in (a,b)$.   Then, $E(\Wz \in A^-_\zd) =  i S^\zt E(\Wz \in
A^+_\zd)$, and the transformation $\zG = i S^\zt X$ has Jacobian
\begin{equation}
\left| 
\frac{d \zG}{d X} \right|
=\exp\left( - \int_0^\zt \zL(S^s X) d s \right) = e^{-\zL_{0,\zt}(X)} ~,
\label{ccordtransf}
\end{equation}
Then, introducing  $\left\langle \mathcal{O} \right\rangle_{\Wz  \in A^+_\zd}$
as the average  of $\mathcal{O}$ according to $\mu$,  under the condition that
$\Wz  \in A^+_\zd$,  and taking  the dissipation  function as  the observable,
$\overline{\mathcal{O}}_{0,\tau}=\Wz$, one may write
\bea
\frac{\mu(E(\Wz \in A^+_\zd))}{\mu(E(\Wz \in A^-_\zd))} 
&=&\frac{ \int_{E(\Wz \in A^+_\zd)} f (\zG) d \zG }{
\int_{E(\Wz \in A^+_\zd)}
f (S^\zt X) e^{-\zL_{0,\zt}(X)} d X } \nonumber \\
&=&\frac{ \int_{E(\Wz \in A^+_\zd)} f (\zG) d \zG }{
\int_{E(\Wz \in A^+_\zd)} e^{-\zW_{0,\zt}(X)} f(X) d X }  
=\left\langle e^{-\zW_{0,\zt}} \right\rangle_{\Wz \in
  A^+_\zd}^{-1}  \ ,
\label{ESFRp}
\eea
{\it i.e.},
\begin{equation} \label{ESFR}
\frac{\mu(E(\Wz \in A^+_\zd))}{\mu(E(\Wz \in A^-_\zd))} =
e^{[A+\ze(\zd,A,\zt)]\zt} \ ,
\end{equation}
{with  $\ze$  an  error  term  due  to the  finiteness  of  $\zd$,  such  that
  $|\ze(\zd,A,\zt)|  \le  \zd$.  We   call  \eref{ESFR}  the  {\em  transient}
  $\zW$-FR.}  The  transient $\zW$-FR refers to  the non-invariant probability
measure $\mu$ of density $f$; it  is remarkable that time reversibility is the
only ingredient of  its derivation.  To obtain the  steady state $\zW$-FR, let
averaging begin at time $t_0$ and consider
\begin{equation}
\frac{\mu(E(\Ft \in A^+_\zd)) }{\mu(E(\Ft \in A^-_\zd)) } \ .
\label{PpoverPp1}
\end{equation}
Taking $t=\zt+2t_0$, the transformation $\zG=iS^t W$ and some algebra yield
\begin{equation}
\frac{\mu(E(\Ft \in A^+_\zd)) }{\mu(E(\Ft \in A^-_\zd)) } = \left\langle \exp \left( - \zW_{0,t} \right) 
\right\rangle_{\overline{\mathcal{O}}_{t_0,t_0+\zt} \in A^+_\zd}^{-1} \ ,
\label{phiratio}
\end{equation}  
and for $\Ft = \Wt$
\begin{equation}
\frac{\mu(E(\Wt \in A^+_\zd))}{\mu(E(\Wt \in A^-_\zd)) }
= e^{\left[ A + \ze(\zd,t_0,A,\zt)\right] \zt}   
\left\langle e^{- \zW_{0,t_0} -  \zW_{t_0+\zt,2t_0+\zt}} 
\right\rangle_{\Wt \in A^+_\zd}^{-1} \ ,
\label{FtWt} 
\end{equation}
where $|\ze(\zd,t_0,A,\zt)| \le \zd$ is due to the finiteness of $A^+_\zd$.

Having fixed  $\zt>0$ and the tolerance $\zd>0$,  we say that $A$  lies in the
domain $\mathcal{D}$ of the steady state $\zW$-FR, if there exists $\hat{t}>0$
such that  $\mu(E(\Wt \in A^+_\zd)) >  0$ and $\mu(E(\Wt  \in A^-_\zd))>0$ for
all $t_0  \ge \hat{t}$. In  other words, $A  \in \mathcal{D}$ if  positive and
negative  fluctuations of  size $A$  have positive  probability in  the steady
state. Using $\mu(E) = \mu_{t_0}(S^{t_0} E)$,  where $E$ is a subset of ${\cal
  M}$, and $\mu_{t_0}$  is the evolved measure up to  time $t_0$, with density
$f_{t_0}$, some algebra yields the $\mathcal{O}$-FR:
\begin{equation}
\frac{\mu_{
    {t_0}}(E(\overline{\mathcal{O}}_{0,\zt} \in A^+_\zd))} {\mu_{
    {t_0}}(E(\overline{\mathcal{O}}_{0,\zt} \in A^-_\zd))} = 
  \frac{\mu(E(\Ft \in A^+_\zd))}{\mu(E(\Ft \in A^-_\zd))} = 
\langle \exp \left(-\Omega_{0,t} \right) \rangle^{-1}_{\Ft \in A^+_\zd} \ .
\label{ESSFT}
\end{equation}
For $\Ft=\Wt$, taking the logarithm and dividing by
$\zt$ produces:
\begin{equation}
\hskip -40pt
\frac{1}{\zt} \ln
\frac{\mu_{t_0}(E(\overline{\zW}_{0,\zt} \in A^+_\zd))}
{\mu_{t_0}(E(\overline{\zW}_{0,\zt} \in A^-_\zd))} = 
A + \ze(\zd,t_0,A,\zt) 
- \frac{1}{\zt} \ln 
\left\langle e^{-\zW_{0,t_0} - \zW_{t_0+\zt,2t_0+\zt}} \right\rangle_{\Wt \in A^+_\zd}
\label{SSESFT}
\end{equation}
If $\mu_{t_0}$  tends to  a steady state  $\mu_\infty$ when $t_0  \to \infty$,
Eq.(\ref{SSESFT}) should change from a statement on the ensemble $f_{t_0}$, to
a statement on the statistics generated  by a single typical trajectory. To be
of practical use,  however, this statement requires that  the logarithm of the
conditional average, divided by $\zt$, $M(A,\zd,t_0,\zt)$ say, be controllable
in  Eq.(\ref{SSESFT}). For  instance,  if  it can  be  made negligible,  e.g.\ 
letting $\zd$  be small and  $\zt$ grow after  the $t_0 \to \infty$  limit has
been taken, as in the case of the $\zL$-FR, one would have the

\vskip 5pt  \noi {\bf Steady State  $\zW$-FR.} {\it For  any tolerance $\zg>0$
and $A \in \mathcal{D}$, there are sufficiently small $\zd > 0$ and large $\zt$,
such that
\begin{equation} 
A - \zg \le 
\frac{1}{\zt} \ln 
\frac{\mu_{\infty}(E(\overline{\zW}_{0,\zt} \in A^+_\zd))}
{\mu_{\infty}(E(\overline{\zW}_{0,\zt} \in A^-_\zd))} \le A + \zg 
\label{SSFTestim}
\end{equation}
holds.
}

\vskip 5pt \noi As in the case of the $\zL$-FR, the domain $\mathcal{D}$ would
be model  dependent, and  its expression could  rest on  non-trivial dynamical
relations \cite{GG-MPEJ}.   This requires some assumption.  Indeed, the growth
of  $t_0$  could  make  $M(A,\zd,t_0,\zt)$  diverge (as  in  properly  devised
examples  \cite{ESR2}).  If $\lim_{t_0  \to  \infty}  | M(A,\zd,t_0,\zt)|$  is
bounded by  some finite  $M(A,\zd,\zt)$, $\lim_{\zt \to  \infty} M(A,\zd,\zt)$
could still exceed the value of  $\zg$.  The first difficulty is simply solved
by  the  observation  that  the  divergence of  $M(A,\zd,t_0,\zt)$  implies  a
divergence of  the left  hand side of  Eq.(\ref{SSESFT}), which in  turn means
that one of  its two probabilities vanish, i.e.\ that  $A \notin \mathcal{D}$. 
If  $\mathcal{D}$ is  empty,  the steady  state  $\zW$-FR is  of no  interest,
because there are no fluctuations in the steady state.

Therefore,  let us  assume  that $A  \in  \mathcal{D}$, and  observe that  the
conservation of probability yields the relation
\begin{equation}
\left\langle e^{-\zW_{0,s}} \right\rangle = 1 ~, \quad \mbox{for every } s \in \zR ~,
\label{normalizat}
\end{equation}
first derived  by Morriss  and Evans (cf.  \cite{EM}, pp.198-202).   Then, one
possibility  that can  be considered  is that  the  $\zW$-autocorrelation time
vanishes. In that case, one can write:
\begin{equation}
1 = \left\langle e^{-\zW_{0,s} -\zW_{s,t} } \right\rangle = 
\left\langle e^{-\zW_{0,s}} \right\rangle \left\langle e^{-\zW_{s,t}} \right\rangle ~, \quad
\left\langle e^{-\zW_{s,t}} \right\rangle =1 ~, \quad \mbox{for all} ~ s, t ~,
\end{equation}
hence
\begin{equation}
\hskip -40pt
\left\langle e^{-\zW_{0,t_0}} \cdot e^{-\zW_{t_0+\zt,2t_0+\zt}} 
\right\rangle_{\Wt\in A^+_\zd} = 
\left\langle e^{-\zW_{0,t_0}} \cdot e^{-\zW_{t_0+\zt,2t_0+\zt}} \right\rangle  
= 1 ~.
\label{condave-full}
\end{equation}
Then, the logarithmic correction  term in (\ref{SSESFT}) identically vanishes
for all $t_0,\zt$, and the \WFR is  verified at all $\zt > 0$. Of course, this
idealized  situation does  not need  to be  realized, but  tests  performed on
molecular dynamics systems \cite{ESR3}  indicate that the typical situation is
not dissimilar from this; typically, there exists a constant $K$, such that
\begin{equation}
0 < \frac{1}{K} \le \left\langle e^{-\zW_{0,t_0} - \zW_{t_0+\zt,2t_0+\zt}} 
\right\rangle_{\Wt \in A^+_\zd} \le K ~.
\label{aveBDD}
\end{equation}
As a  matter of fact,  the de-correlation or  Maxwell time, $t_M$,  expresses a
physical property  of the system, thus it  does not depend on  $t_0$ or $\zt$,
and depends only mildly on the external field [usually, $t_M(F_e) = t_M(0) + O
(F_e^2)$]. Its order of  magnitude is that of the mean free  time. If $\zt \gg
t_M$, the  boundary terms $\zW_{t_0-t_M,t_0}$  and $\zW_{t_0+\zt,t_0+\zt+t_M}$
are typically  small compared to $\zW_{t_0,t_0+\zt}$,  unless some singularity
of $\zW$  occurs within $(t_0-t_M,t_0)$  or $(t_0+\zt,t_0+\zt+t_M)$.  However,
similar   events  may   equally   occur  in   the   intervals  $(0,t_0)$   and
$(t_0+\zt,2t_0+\zt)$,          hence          $\zW_{t_0-t_M,t_0}$          and
$\zW_{t_0+\zt,t_0+\zt+t_M}$  are expected  to  contribute only  a fraction  of
order $O(t_M/\zt)$  to the  arguments of the  exponentials in  the conditional
average.  Therefore, one can write
\begin{eqnarray}
\hskip -50pt\left\langle
e^{-\zW_{0,t_0}} \cdot e^{- \zW_{t_0+\zt,2t_0+\zt}} 
\right\rangle_{\overline{\zW}_{t_0,t_0+\zt} \in A_\zd^+} &\approx& 
\left\langle e^{-\zW_{0,t_0-t_M}} \cdot e^{- \zW_{t_0+\zt+t_M,2t_0+\zt}}
\right\rangle_{\overline{\zW}_{t_0,t_0+\zt} \in A_\zd^+} \nonumber \\
&\approx& \left\langle e^{-\zW_{0,t_0-t_M}} \cdot e^{- \zW_{t_0+\zt+t_M,2t_0+\zt}}
\right\rangle \nonumber \\
&\approx& \left\langle e^{-\zW_{0,t_0+t_M}} \right\rangle
\left\langle e^{- \zW_{t_0+\zt+t_M,2t_0+\zt}} \right\rangle = O(1) \ ,
\label{newargdeco}
\end{eqnarray} 
with an accuracy which improves with growing $t_0$ and $\zt$, because $t_M$ is
fixed.   If  these  scenarios  are  realized,  Eq.(\ref{aveBDD})  follows  and
$M(A,\zd,t_0,\zt)$ vanishes  as $1/\zt$, with a characteristic  scale of order
$O(t_M)$.  In  summary, the  steady state $\zW$-FR  holds under  the following
conditions.

\vskip 5pt
\noi 
{\bf Conditions: }\\
{\em {\bf 1.} the dynamics is time reversal invariant. \\
\noi
{\bf 2.} $\mu_t$ tends to $\mu_\infty$ for $t \to \infty$. \\
\noi
{\bf 3.} Eq.(\ref{aveBDD}) is satisfied with $K > 0$,
for $A \in \mathcal{D}$, if $\zt$ and $t_0$ are sufficiently
larger than $t_M$.
}

\vskip 5pt  \noi Condition  (\ref{aveBDD}) can actually  be weakened,  but the
decay of the $\zW$-autocorrelations  characterizes the convergence to a steady
state,   and   is  very   widely   verified.   Therefore,   the  validity   of
Eq.(\ref{aveBDD}), and not  a weaker condition, explains why  the steady state
$\zW$-FR  holds  for the  particle  systems  so  far investigated.  The  above
derivation of the steady state $\zW$-FR, under Conditions 1, 2 and 3, will not
only answer the  physics questions, but will also  be mathematically rigorous,
if it  will be proven  that one (possibly physically  uninteresting) dynamical
system satisfies them.

Various other relations can now be obtained \cite{ESR2}. For instance, any odd
$\mathcal{O}$, any $\zd>0$, any $t_{0}$ and any $\zt$ yield
\begin{equation} 
\langle
\exp \left( -\Omega_{0,t} \right) \rangle_{\Ft \in (-\zd,\zd)}
= \frac{\mu_{t_0}(E(\overline{\mathcal{O}}_{0,\zt} \in (-\zd,\zd)))}
{\mu_{t_0}(E(\overline{\mathcal{O}}_{0,\zt} \in (-\zd,\zd)))}
= 1 \ , \label{inter} 
\end{equation}
which,  in  the  $\zd\to\infty$  limit, produces  the  normalization  property
(\ref{normalizat}).   The  Dissipation  relation
\begin{equation}
\langle  \mathcal{O} \rangle_t = \int_0^t d s \langle  \zW(0) \mathcal{O}(s)
\rangle  
\end{equation}
is  another  direct consequence  of  the  approach  followed in  this  section
\cite{ESdissir}.

\subsection{Green-Kubo relations}
A consistency check of the present theory is afforded by the derivation of the
Green-Kubo  relations  based  on  the $\zW$-FR  \cite{ESR}.  Differently  from
Ref.\cite{GG96}, which deals  with time-asymptotic quantities, this derivation
stresses  the  role of  the  physical  time scales.  To  be  concrete, take  a
Nos\'e-Hoover thermostatted  system, whose  equilibrium state is  the extended
canonical density
{
\begin{equation}
f_c(x,\za) = 
\frac{e^{-\zb \left(H_0+Q \za^2 /2 \right)}}{\int d \za ~ d x~~ 
e^{-\zb \left(H_0+Q \za^2 /2 \right)}} \ ,
\end{equation}
where $Q  = 2  K_0 \zt^2$ and  $H_0$ is  the internal energy  \cite{EM}.} This
yields
\begin{equation}
f_c(\za) = \int d x f_c(x,\za) = \sqrt{\frac{\zb Q}{2 \pi}} ~ \exp \left[- \zb Q \za^2 /2 \right]
\end{equation}
Therefore,  the distribution of  $\za_{0,t}$ is  Gaussian in  equilibrium, and
near equilibrium it can be assumed  to remain such, around its mean, for large
$t$ (CLT).  To use the FR together  with the CLT, the values $A$ and $-A$ must
be a small  number of standard deviations away from  $\langle \zW \rangle$. In
\cite{SE2000} it was proven that
\begin{displaymath}
t \zs_{J_t}(F_e) = 2 L(F_e) k_{_B} T/V + O((F_e)^2/t N) ~,
\end{displaymath}
where
\begin{displaymath}
L(F_e) = \zb V \int_0^\infty d t \langle (J(t)-\langle J \rangle_{F_e})
((J(0)-\langle J \rangle_{F_e}) \rangle_{F_e} ~,
\end{displaymath}
$F_e$ is  the external field,  $\langle\cdot\rangle_{F_e}$ is the  phase space
average  at  field  $F_e$  and  $L(0)  =  \lim_{F_e  \to  0}  L(F_e)$  is  the
corresponding linear  transport coefficient.  When $t$ grows,  $A=0$ gets more
and  more  standard deviations  away  from  $\langle  \zW \rangle$,  which  is
$O(F_e^2)$, for small $F_e$, while  the standard deviation tends to a positive
constant,  since  that  of  $\za$  tends  to  $1/\sqrt{\zb  Q}$.   Assume  for
simplicity  that the  variance of  $\zW_{0,t}(F_e)$ is  monotonic in  $F_e$ at
fixed $t$, and in $t$ at fixed $F_e$.  Then, there is $t_\zs(F_e,A)$ such that
the variance is sufficiently large when $t < t_\zs(F_e,A)$.  At the same time,
$t$ has to be larger than a  given $t_\zd(F_e,A)$ for the steady state \WFR to
apply  to the  values $A$  and  $-A$, with  accuracy $\zd$.  Assume that  also
$t_\zd(F_e,A)$ is monotonic in $F_e$.  To derive the Green-Kubo relations, one
then  needs $  t_\zd(F_e,A) <  t <  t_\zs(F_e,A)$ for  $F_e \to  0$,  which is
possible because the  distribution tends to a Gaussian  centered in zero, when
$F_e$ tends to zero and $t$ is fixed. The result is:
\begin{equation}
\langle \zW \rangle = \frac{1}{2} \zs^2(\zW) \quad \mbox{or } ~~
L(0) = \lim_{F_e \to 0} \frac{\langle J \rangle_{F_e}}{F_e} = \zb V 
\int_0^\infty d t ~ \langle J(0) J(t) \rangle_{F_e=0} \ .
\end{equation}

\subsection{Discussion}
The analysis  of this  section shows  that the steady  state $\zW$-FR  and its
consequences  can  be obtained  only  from  time  reversibility and  from  the
$\zW$-autocorrelation decay.  These are the physical mechanisms underlying the
validity of the steady state $\zW$-FR and, indeed, they correctly identify the
relevant time scales.  From a purely mathematical point of  view, the decay of
the $\zW$-autocorrelation could be relaxed,\footnote{It suffices that the $t_0
  \to \infty$ limit of the conditional average of Eq.(\ref{SSESFT}) grows less
  than exponentially  fast, with $\zt$, or  that its exponential  growth has a
  rate smaller  than $\zd$.}  but  is needed for  the convergence to  a steady
state.  Therefore, the  systems that verify the steady state  \WFR do not need
to have any  (even approximate) Anosov structure. At the  same time, the above
analysis  does not identify  the class  of dynamical  systems which  enjoy the
required  $\zW$-autocorrelation decay, as  needed to  make rigorous  the above
derivation of  the steady  state $\zW$-FR. However,  this does not  impair our
understanding of the physics of  the steady state $\zW$-FR, while the explicit
construction of artificial models  verifying (\ref{aveBDD}) is not necessarily
physically revealing.

How  can the  above analysis  be reconciled  with axiom  C systems,  and their
modified $\zL$-FR, Eq.(\ref{axiomCFR}),  introduced in Refs.\cite{BGG,axiomC}? 
Axiom C  systems, indeed, enjoy a  strong decay of  correlations and, although
there  are  no particle  systems  known  to be  of  their  kind,  they can  be
abstractly conceived. Furthermore, modified FRs have been observed to hold for
some observables,  in particular dynamical  systems \cite{GRS}. The  answer is
that the  decay of correlations  of axiom C  systems does not imply  the decay
required  here: the first  concerns all  observables, and  is referred  to the
invariant  measure; the second  concerns only  $\zW$, and  is referred  to the
initial measure \cite{ESR2}. Therefore, certain  dynamics may enjoy a decay of
correlations with respect  to $\mu_\infty$, while they do  not with respect to
$\mu$, and  no contradiction arises. What  happens, in general, is  not known. 
In Ref.\cite{ESR2},  Appendix 2, the behaviour of  $M(A,\zd,t_0,\zt)$ has been
explicitly computed  for the  isokinetic particle in  free space,  proposed in
Ref.\cite{CG99}. It was found that $M$ diverges, hence that it does not verify
(\ref{aveBDD}), and  that the $\zW$-FR does  not apply, in  agreement with the
fact that  the steady state of that  system has no fluctuations.  The study of
more general cases is desired. It is also desired that the physical meaning of
condition (\ref{aveBDD})  be better understood. Indeed,  close to equilibrium,
the decay of  correlations with respect to the  equilibrium measure amounts to
the standard condition,  required by the Green-Kubo theory,  for the existence
of  the  transport  coefficients.  Far   from  equilibrium,  it  needs  to  be
understood.

Given  that  Eq.(\ref{SSESFT})  is   an  exact  result,  various  mathematical
questions arise.   Can one find dynamical  systems and functions  of phase for
which Eq.(\ref{aveBDD})  holds? How does $M(A,\zd,t_0,\zt)$ behave  in axiom C
systems, in general? What happens with the steady state fluctuations of $\zW$,
if $\zW$ is bounded?

\section{Work relations: Jarzynski and Crooks}
\label{sec:JE}

Consider a finite  particle system, in equilibrium with  a much larger system,
which  constitutes a heat  bath at  temperature $T$.  Assume that  the overall
system is  described by a Hamiltonian  of the form  \be \mathcal{H}(\zG;\zl) =
H(x;\zl) +  H_E(y) + h_i(x,y) \ee where  $x$ and $y$ denote  the positions and
momenta of the  particles of, respectively, the system of  interest and of the
bath, $h_i$ represents  the interaction between system and  bath, and $\zl$ is
an  externally controllable  parameter. This  system can  be driven  away from
equilibrium, performing work $W$ on it, by acting on $\zl$. Let $\zl(0)=A$ and
$\zl(\zt)=B$ be the  initial and final values of $\zl$,  for a given evolution
protocol $\zl(t)$.  Suppose the process is  repeated very many  times to build
the statistics of the  work done, varying $\zl$ from $A$ to  $B$ always in the
same manner. Let  $\rho$ be the PDF of the externally  performed work. This is
not the thermodynamic work done on the system, if the process is not performed
quasi  statically \cite{CohMauz},  but is  always a  measurable  quantity. The
Jarzynski Equality predicts that \cite{CJ}:
\begin{equation}
\left\langle e^{-\zb W} \right\rangle_{A \to B} = 
\int d W ~ \rho(W) e^{- \zb W} = e^{-\zb \left[ F(B) - F(A) \right]}
\label{JR}
\end{equation}
where $\zb  = 1  / k_{_B} T$,  and $\left[  F(B) - F(A)  \right]$ is  the free
energy difference between the initial  equilibrium state, with $\zl=A$ and the
equilibrium  state  which is  eventually  reached  for  $\zl=B$.  The  average
$\langle e^{-\zb W}  \rangle_{A \to B}$ is the average over  all works done in
varying  $\zl$ from  $A$  to $B$.   While  the process  always  begins in  the
equilibrium state corresponding to $\zl=A$, the  system does not need to be in
equilibrium when $\zl$ reaches the  value $B$.  However, the equilibrium state
with  $\zl=B$ exists and  is unique,  hence $F(B)$  is well  defined. Equation
(\ref{JR}) is supposed to hold  whichever protocol one follows to change $\zl$
from  $A$  to  $B$,  hence   also  arbitrarily  far  from  equilibrium  (large
$\dot{\zl}$); therefore the presence  of the equilibrium quantities $F(A)$ and
$F(B)$ in Eq.(\ref{JR}) is remarkable.   From the thermodynamic point of view,
one observes that the externally measured  work does not need to coincide with
the internal  work (which would not  differ from experiment  to experiment, if
performed quasistatically).   From an operational  point of view, it  does not
matter whether the  system is in local equilibrium or  not: certain forces are
applied, certain  motions are  registered, hence certain  works are  recorded. 
The Jarzynski equality is a transient relation and, similarly to the transient
$\zW$-FR, rests on minimal conditions  on the microscopic dynamics. It is also
consistent with the second law of thermodynamics, since it yields
\begin{equation}
\left\langle \zb W \right\rangle_{A \to B} \ge \zb \left[ F(B) - F(A) \right]
\end{equation}
because  $\ln  \left\langle  \Phi  \right\rangle  \ge  \left\langle  \ln  \Phi
\right\rangle$ for positive observables $\Phi$.  

Similarly, computing  the ratio of the  probability that the work  done in the
forward transformation is  $W$, to the probability that it is  $-W$ in the $B$
to  $A$  transformation, with  reversed  protocol  $-\dot{\zl}$, produces  the
Crooks Relation \cite{GK}:
\begin{equation}
\frac{P_{A \to B} (W = a)}{P_{B \to A} (W = -a)} = 
e^{-\zb \left[ F(B) - F(A) \right]} ~
e^a
\end{equation}
The Crooks Relation, leads to the Jarzynski Equality, by a simple integration:
\begin{eqnarray}
\hskip -50pt\left\langle e^{-\zb W} \right\rangle_{A \to B} &=& 
\int P_{A \to B} (W = a) e^{-a} d a = e^{-\zb \left[ F(B) - F(A) \right]}
\int P_{B \to A} (W = -a) d a \\
&=& e^{-\zb \left[ F(B) - F(A) \right]}
\end{eqnarray}
These  results  and the  $\zW$-FR  are connected.   In  the  first place,  the
transient \WFR may  be applied to the protocols of  the Jarzynski Equality and
of  the Crooks  Relation,  \cite{DJE03}.  Then,  let  $f_A$ and  $f_B$ be  the
canonical distributions at same inverse temperature $\zb$ for the Hamiltonians
$H_A$  and $H_B$  of the  equilibrium states  $A$ and  $B$  respectively.  The
corresponding Helmholtz free energies {are $F_i  = - k_{_B} T \ln \int d x
  ~ \exp[-H_i(x)/k_{_B}  T]$ for} $i=A,B$.  For simplicity, let $\zl$  go from
$A$ to $B$ in a time $\zt$, with rate $\dot{\zl} = 1/\zt$, and from $B$ to $A$
with rate  $\dot{\zl} =  -1/\zt$.  Correspondingly, a  thermostatted evolution
may be defined by
\begin{eqnarray}
&&\hskip -40pt \dot{\zl} = \pm \frac{1}{\zt} \nonumber \\ 
&&\hskip -40pt \dot{\q} = \frac{\partial H(\q,\p;\zl)}{\partial \p} 
~; \quad \dot{\p} = - \frac{\partial H(\q,\p;\zl)}{\partial \q} - S_i \za(x) \p ~; 
\quad  S_i = \left\{ \begin{array}{ll} 1 , & i=1,..., N_w \\ 0 , & i > N_w
\end{array}
\right.
\end{eqnarray}
where the thermostat  acts only on $N_w$ particles (the  walls of the system),
to  fix their  kinetic  temperature and  mimic  a heat  bath.  Then, the  work
performed by  the external  forces is given  by $\zb W  = \zb  [ H_B -  H_A] -
\zL_{0,\zt}$.  If $A$ and $B$ are the same equilibrium state, the \WFR applies
directly, and the  Jarzynski equality is an immediate  consequence of the \WFR
because the \WFR implies
\begin{equation}
\left\langle e^{-\zb W} \right\rangle = \int P(W) e^{-\zb W} d W \int d W P(-W) = 1 ~.
\end{equation}

The $\zW$-FR, the Jarzynski Equality and  the Crooks Relation do not have same
range  of  applicability, the  Crooks  Relation  being  the most  general  for
canonical  ensembles   \cite{DJE03}.   It  is  remarkable   how  they  connect
equilibrium to  nonequilibrium properties of physical  systems; their interest
is bound to  grow with our understanding of  microscopic systems, particularly
in nanotechnology and biophysics \cite{bustamante}. One reason for considering
NEMD models  in this  context, is that  they afford  heat baths which  are not
affected  by the transformation  processes during  which work  is done  on the
system  of  interest.   Differently,  finite Hamiltonian  reservoirs  are  not
guaranteed to  be as isothermal as  required. Although the effect  of the work
done may be negligible on averages, if the reservoirs are large, its influence
on  fluctuations could  be sizeable,  especially in  particular circumstances,
like  around phase  transitions.  Therefore,  that different  approaches agree
where appropriate, strengthens all results.

\section{Stochastic systems and the Van Zon - Cohen extended FR}
\label{sec:vzc}

The  first stochastic  FR  was derived  by  Kurchan, who  obtained a  modified
detailed balance property  for Langevin processes of finite  systems, and a FR
for the entropy  production, under a few assumptions,  like the boundedness of
the potentials \cite{Kurchan}.  In 1999, Lebowitz and Spohn \cite{LS} extended
Kurchan's results to generic Markov processes: under the assumption that local
detailed balance is attained, they  showed that the Gibbs entropy variation is
related to the action functional that  satisfies the FR. This suggests that in
Markov processes the Gibbs entropy variation plays the role of the phase space
contraction.  In  Ref.\cite{CM99}, Maes  obtained a large  deviation principle
for discrete  space-time Gibbs measures, leading to  a FR for a  kind of Gibbs
entropy variation in time discrete lattice systems.  These results can be seen
as a generalization of the  $\Lambda$-FR and of its stochastic versions, since
stochastic   dynamics  and  thermostatted   systems  satisfying   the  chaotic
hypothesis are examples of systems with space-time Gibbs measures.

In 2002, Farago  pointed out that singularities may  cause difficulties in the
conventional use of stochastic FRs  \cite{Farago}. In the same year, Wang {\em
  et.al.\ } reported the experimental verification of an integrated version of
the \WFR  for colloidal particles dragged  through water, by  a moving optical
trap  \cite{WSMSE}.  This experiment  may  be  modeled  through an  overdamped
Langevin process,  describing a  Brownian particle, dragged  in a liquid  by a
moving    harmonic    potential    with    a   constant    velocity    ${v}^*$
\cite{Farago,VZCa,VZCb,mazonka,narayan,Maesrecent}:
\begin{equation} \label {eq:brownian}
\deriv{x(t)}{t} = -(x(t) - x^*(t)) + \zeta(t) ~.
\end{equation}
Here $x(t)$ is the  position of the particle at time $t$,  $x^*(t) = v^*t$ the
position of  the minimum of  the potential, $\zeta(t)$  is a white  noise term
representing the fluctuating force the fluid exerts on the particle, and $k_BT
= 1$.  Then, the work done in a time $\zt$ is
\begin{equation} \label{eq:vzc-work}
W_\tau = v^* \int_0^\tau [-(x(t) - x^*(t)] dt \ .
\end{equation}
Analyzing the  results of \cite{WSMSE}  from this point  of view, Van  Zon and
Cohen \cite{VZCa,VZCb} considered separately the  
{\em dissipated} energy, or the heat $Q_\tau$, and the potential energy of the 
Brownian particle $\Delta U_\tau$, and
\begin{equation} \label{eq:e-balance}
W_\tau = Q_\tau + \Delta U_\tau \ .
\end{equation}
In  \cite{vzc},  Van  Zon  and   Cohen  showed  that,  in  a  comoving  frame,
Eq.~(\ref{eq:brownian}) reduces  to a standard  Ornstein-Uhlenbeck process and
thus,  the  stationary  probability  distribution  and  Green's  function  are
Gaussian in  the particle's position.  Since  the total work is  linear in the
particle's position, $W_\tau$  is Gaussian as well.  {Because  of this and
  of  Eq.~(\ref{eq:vzc-work}),  the  variance  of  transient  fluctuations  of
  $W_\tau$ equals $2 \langle W_\zt \rangle$,} and the total work satisfies the
transient FR.  In the  $\tau\rightarrow\infty$ limit, the variance of $W_\tau$
remains twice its mean, hence the total work satisfies the steady state FR.

Van Zon and Cohen clarified  that the experiment of \cite{WSMSE} concerned the
total  work, and  that  the PDF  of  the potential  energy  is exponential  at
equilibrium,  P($\Delta U)  \sim \exp(\Delta  U)$, and  is expected  to remain
exponential away from equilibrium.  Therefore, while the small fluctuations of
heat  are  expected to  coincide  with  those of  the  total  work, since  the
contribution  of the  potential energy  is only  $\mathcal{O}(1)$,  large heat
fluctuations are more likely to be due to a large fluctuation of the potential
energy.

To  summarize  the  derivations  of  Ref.\cite{VZCb},  consider  the  harmonic
potential    $V(x,t)     \equiv    \frac{1}{2}|x(t)    -     x^*(t)|^2$    in
Eq.~(\ref{eq:e-balance}).   Then  the  heat   $Q_\tau$  is  nonlinear  in  the
particle's  position,  hence its  PDF  needs  not  be Gaussian.   Its  Fourier
transform is
\begin{equation} \label{eq:FT-Q}
\hat{P}_\tau(q) \equiv \int_{-\infty}^\infty dQ_\tau e^{iqQ_\tau}
P_\tau(Q_\tau) \ .
\end{equation}
Writing  $P_\tau(Q_\tau)$ in  terms  of  the joint  distribution  of the  work
$W_\tau$ and of the positions $x(0), x(\tau)$, one obtains
\begin{equation} \label{eq:vzc-fourier}
\hat{P}_\tau(q) =
\frac{\exp\left\{w\left(i-q\right)\left(\tau -
      \frac{2q^2(1-e^{-\tau})^2}{1 + (1-e^{-2\tau})q^2}\right)\right\}}{[1 +
  (1-e^{-2\tau})q^2]^{3/2}} \ ,
\end{equation}
where $w=\langle W_\zt \rangle / \zt$ is  the rate of work done in the system,
and  $\langle \cdot \rangle$  is the  steady state  average. Anti-transforming
$\hat{P}_\tau(q)$, one considers the heat fluctuation function
\begin{equation} \label{eq:Q-fluc}
f_\tau(p) =
\frac{1}{w\tau}\ln\left[\frac{P_\tau(pw\tau)}{P_\tau(-pw\tau)}\right] \ ,
\end{equation}
where  $p=Q_\tau/\langle Q_\tau\rangle$ and  $\langle Q_\tau\rangle  = \langle
W_\tau\rangle -  \langle \Delta  U_\tau\rangle = w\tau$,  since $\langle\Delta
U_\tau\rangle = 0$,  in the steady state.  To  obtain an asymptotic analytical
expression of Eq.~(\ref{eq:Q-fluc}), consider the quantity
\begin{equation} \label{eq:e}
e(\lambda) \equiv \lim_{\tau\rightarrow\infty}-\frac{1}{w\tau}\langle
e^{-\lambda Q_\tau} \rangle \ ,
\end{equation}
and, for large $\tau$, 
\begin{equation} \label{eq:Pq}
P_\tau(Q_\tau) \sim e^{-w\tau\hat{e}(Q_\tau/w\tau)} \ ,
\end{equation}
where  $\hat{e}(p)= \max_{\{\lambda\}}  \  [e(\lambda) -  \lambda  p]$ is  the
Legendre transform of $e(\lambda)$.  As  Lebowitz and Spohn proved for a class
of stochastic models \cite{LS}, if the relation
\begin{equation} \label{eq:LS}
e(\lambda) = e(1-\lambda)
\end{equation}
is    satisfied,   then    the   conventional    steady   state    FR   holds:
$\lim_{\tau\rightarrow\infty}f_\tau(p)   =  p$   (cf.\  Eqs.~(\ref{eq:Q-fluc},
\ref{eq:Pq},   \ref{eq:LS})).   Analytically   continuing   $\hat{P}_\tau$  to
imaginary  arguments,   one  gets  $\langle  e^{-\lambda   Q_\tau}  \rangle  =
\hat{P}_\tau(i\lambda)$, i.e.\ 
\begin{equation} \label{eq:e-sol}
\langle e^{-\lambda Q_\tau}\rangle =
\frac{\exp\left[-w\lambda(1-\lambda)\left\{\tau+\frac{2\lambda^2(1-e^{-\tau})^2}{1-(1-e^{-2\tau})\lambda^2}\right\}\right]}{[1-(1-e^{-2\tau})\lambda^2]^{3/2}} ~,
\end{equation}
which  is   singular  for  $\lambda=\pm   (1-e^{-2\tau})^{-1/2}$.   {Using
  Eqs.~(\ref{eq:e})    and    (\ref{eq:e-sol})    and   taking    the    limit
  $\tau\rightarrow\infty$,   the   singularities   move   to  $\pm   1$,   and
  Eq.~\eref{eq:LS}  is  satisfied for  $|\lambda|<1$.  For $|\lambda|>1$,  the
  integral in  Eq.~(\ref{eq:FT-Q}) diverges, because of  the exponential tails
  of    $P_\tau(Q_\tau)$.}    Thus,    substituting   in    $\hat{e}(p)$   and
Eq.~(\ref{eq:Pq}), one obtains
\begin{equation} \label{eq:VZCFT}
\hskip -30pt
\lim_{\tau\rightarrow\infty} f_\tau(p) = \left\{
\begin{array}{ll}
p & \textrm{for} \ 0 \le p < 1\\
p-(p-1)^2/4 & \textrm{for} \ 1 \le p < 3\\
2 & \textrm{for} \ p \ge 3\\
\end{array}
\right. \quad \mbox{where } ~~ f_\tau(-p)=-f_\tau(p) ~,
\end{equation}
i.e.\ the  fluctuations of heat  smaller than $\langle  Q_\tau\rangle$ satisfy
the conventional  FR, like  those of $W_\zt$,  while larger  heat fluctuations
satisfy the modified relation \eref{eq:VZCFT}.

These results do not contradict  those of Ref.\cite{LS}, because $Q_\zt$ lives
in an  infinite state space,  due to the  unboundness of the  potential, while
Ref.\cite{LS}  only  concerns  finite  state spaces.   Therefore,  $Q_\zt$  is
affected by boundary terms which cannot be neglected and which distinguish its
behaviour  from that  of $W_\zt$  \cite{PRV}.  As  discussed in  Section  4, a
similar   phenomenon    may   concern   $\zL$,    in   deterministic   systems
\cite{ESR,BGGZ}.  Therefore, one  may argue that $\zL$ plays  the role of heat
\cite{EvansMS}.   Baiesi et  al.  generalize  the results  of \cite{VZCa,VZCb}
considering a  Langevin process with general confinement  potential and motion
of the minimum of the potential, $x^*$ \cite{Maesrecent}.  They find necessary
conditions on  the potential $V$ and  on its motion $x^*(t)$,  for $W_\tau$ to
satisfy  the  steady  state FR,  namely:  $a$)  $x^*$  must  be even  in  time
($x^*(t)=x^*(-t)$), or $b$) it must be odd ($x^*(t)=-x^*(-t)$) and $V$ must be
even  in space  ($V(x,t)=V(-x,t)$).   Under these  conditions,  they obtain  a
generalization  of  Eq.~\eref{eq:VZCFT}  for  the fluctuations  of  heat.   In
particular, numerical  test shows that  for $x^*$ moving at  constant velocity
and  non-symmetric  $V$'s, $W_\tau$  does  not  satisfy  the steady  state  FR
\cite{Maesrecent}.  Similar observations are reported in \cite{dhar}.

The question of  the validity of the extended FR  of \cite{VZCa,VZCb} has been
addressed  in other  papers, like  Ref.\cite{PRV},  where the  extended FR  is
verified on a granular  system. Differently, Ref.\cite{germans} shows that the
extended FR does not hold  in the partially asymmetric zero-range process with
open  boundaries.   Various  other   studies  have  recently  dealt  with  the
statistical  properties of  Brownian  particles and  Langevin processes,  like
Refs.\cite{PO98,HS01,seifert05,seifert06,Kurchanlast,visco,imparatoa,imparatob}.

\section{Temporal asymmetry of fluctuations}
\label{sec:JL}

References  \cite{bsgj01a,bsgj01b,BDSJLcurrent}   propose  extensions  of  the
Onsager-Machlup  theory  \cite{OM53a,OM53b}   to  the  large  fluctuations  of
physical systems  in nonequilibrium steady  states, from which  stochastic FRs
can  be  obtained.   For  density-like  observables  of  stochastic  processes
describing  nonequilibrium  systems in  local  thermodynamic equilibrium,  the
theory     predicts    temporal     asymmetries    in     the    corresponding
fluctuation-relaxation paths (FRPs).

For  a  class  of  stochastic  lattice gases,  which  admit  the  hydrodynamic
description
\begin{equation}
\partial_{t}\varrho = \nabla \cdot
\left[\frac{1}{2} D\left(\varrho \right)\nabla \varrho \right] \equiv \mathcal{D}\left(\varrho
\right) ~, \quad \quad \varrho = \varrho(u,t) ~,
\label{hydro}
\end{equation}
where  $\varrho$  is  the  vector  of  macroscopic  observables,  $u$  is  the
macroscopic space  variable, $t$ is the  macroscopic time, $D$  is the Onsager
diffusion  matrix, let  $\hat{\varrho}$ be  the steady  state, with  the given
boundary  conditions.   Then,   Refs.\cite{bsgj01a,bsgj01b}  proves  that  the
spontaneous fluctuations out of a steady state, are governed by a certain {\em
  adjoint} hydrodynamic equation:
\begin{equation}
  \partial_{t}\varrho =\mathcal{D}^{*}\left(\varrho \right),
\label{adjointH}
\end{equation}
with same boundary  conditions. This is supposed to  hold much more generally;
namely, whenever the following holds \cite{bsgj01a,bsgj01b}:

\vskip 2pt  \noindent {\bf Assumptions:}  1) {\it The mesoscopic  evolution is
  given by a  Markov process $X_t$, which represents  the configuration of the
  system  at time  $t$.  The nonequilibrium  steady  state is  described by  a
  probability measure $P_{st}$ over the trajectories of $X_t$;}

\noindent 2) {\it the fields $\varrho$ obeying Eq.(\ref{hydro}) constitute 
  the  local thermodynamic  variables, and  the steady  state under  the given
  boundary conditions is unique;}

\noindent 3)
{\it Denoting by $\theta$ the time inversion operator defined by $\theta X_t =
  X_{-t}$, the probability measure $P_{st}^*$, describing the evolution of the
  time reversed process $X_t^*$, and $P_{st}$ are related by
\begin{equation}
P^*_{st}(X_t^*=\varphi_t, t\in[t_1,t_2]) = P_{st}(X_t=\varphi_{-t},t\in[-t_2,-t_1]).
\label{Pst*}
\end{equation}
If $L$  is the generator  of $X_t$, the  adjoint dynamics is generated  by the
adjoint (with  respect to the  invariant measure $\mu$) operator  $L^*$, which
admits the {\em adjoint} hydrodynamics (\ref{adjointH}) };

\noindent
4) {\it The measure $P_{st}$ admits a large deviation principle describing the
  fluctuations of $\varrho$. }

\vskip  5pt \noi  This is  the mesoscopic  evolution, which  is a  reduced, or
coarse grained,  description of underlying deterministic  dynamics. It implies
that spontaneous macroscopic fluctuations out of a nonequilibrium steady state
most  likely follow  a trajectory  which  is the  {\em time  reversal} of  the
relaxation path, according to the adjoint hydrodynamics, i.e.\ 
\begin{equation}
\partial_t \varrho = \mathcal{D}^*(\varrho) = \mathcal{D}(\varrho) - 2 \mathcal{A} ~,
\label{Ddiff}
\end{equation}
where $\mathcal{D}$ can be decomposed as
\begin{equation}
\mathcal{D}(\varrho) = \frac{1}{2} \nabla \cdot \left( \chi(\varrho) \nabla
\frac{\delta \mathcal{S}}{\delta \varrho} \right) + \mathcal{A} ~
\label{decompose}
\end{equation}
{and $\mathcal{A}$ is a vector field orthogonal to the thermodynamic force
  $\delta  \mathcal{S}   /  \delta  \varrho$.  Thus   $\mathcal{A}$  does  not
  contribute to the entropy production.}   In the limit of small fluctuations,
and small differences in the  chemical potentials at the boundaries, Onsager's
theory is recovered, because $\mathcal{A}$ is a higher order term.

Although  the stochastic  behaviour should  be  a coarser  description of  the
deterministic   one,   at   present    the   gap   between   the   theory   of
\cite{bsgj01a,bsgj01b} and  the behaviour of  systems such as the  NEMD models
does not seem  to be bridgeable in rigorous terms.   Thus, one wonders whether
the predictions of Refs.\cite{bsgj01a,bsgj01b,BDSJLcurrent} may be verified in
{\em  reversible}  deterministic particle  systems.   In  particular, are  the
corresponding FRPs  {\em asymmetric} in time?  This is important,  in order to
understand  how common  the asymmetric  behaviour might  be  in nonequilibrium
phenomena. Also,  the temporal asymmetry  of fluctuations has some  bearing on
the  question of  how macroscopic  irreversibility relates  to  the reversible
microscopic  dynamics  \cite{lc97},  a  question  which  cannot  be  addressed
investigating intrinsically  irreversible stochastic systems.   Therefore, the
stochastic  approach needs  to be  complemented by  the deterministic  one. In
Ref.\cite{GR04},  no temporal  asymmetry  was detected  in the  nonequilibrium
Lorentz   gas;    while   in   Refs.\cite{GRV-jonaa,GRV-jonab,PSR},   temporal
asymmetries  were  found  in the  FRPs  of  the  nonequilibrium FPU  model  of
\cite{LLP1}, and of the SLLOD model.

The  origin  of  the   temporal  asymmetry  may  be  heuristically  understood
considering the macroscopic deterministic (irreversible) dynamics described by
\begin{equation}
\dot{\varrho} = {\cal D}(\varrho) ~,
\label{xdotb}
\end{equation}
on ${\cal M} \subset \zR^n$, where ${\cal  D}$ is a vector field with a unique
attracting      fixed     point      $\hat{\varrho}     \in      {\cal     M}$
\cite{GR04,GRV-jonaa,GRV-jonab}.  This   is  compatible  with  microscopically
reversible dynamics, in which case $\hat{\varrho}$ has a repelling counterpart
$\tilde{\varrho}$.  The  $n$ components of $\varrho$ may  represent the values
taken by  a scalar thermodynamic  observable on the  $n$ different sites  of a
spatially  discrete   system.   Let  the   local  mesoscopic  dynamics   be  a
perturbation  of  Eq.(\ref{xdotb}),  with   a  Gaussian  noise  of  covariance
$\left\langle \xi _{i}\left( t\right) \xi _{j}\left( t'\right) \right\rangle =
K_{ij}\delta \left( t-t'\right)$ and mean $\left\langle \xi(t) \right\rangle =
0$, where \( K \) is a symmetric, positive definite matrix:
\begin{equation}
\dot{\varrho} = {\cal D}(\varrho) + \xi ~.
\label{stoc-pertu}
\end{equation}
This    allows    different    evolutions    between   one    initial    state
\(\varrho_{i}=\varrho    \left(   t_{i}\right)\)    and   one    later   state
\(\varrho_{f}=\varrho \left(  t_{f}\right)\).  The different  paths connecting
\( \varrho_{i} \) to \( \varrho _{f} \) occur with different probabilities, $P
\propto  \exp \left(  -I_{\rm  path} \right)$,  and,  Eq.(\ref{xdotb}) can  be
obtained from the minimization of the terms
\begin{equation}
I_{\rm path}\left( \varrho \right)  \equiv \frac{1}{2}\int _{t_{i}}^{t_{f}}\left\langle
\dot{\varrho}-\mathcal{D}\, ,\, \dot{\varrho}-\mathcal{D}\right\rangle \,
\mathrm{d}t~,
\end{equation}
where  $\langle x  ,  y  \rangle =  x^T  K^{-1} y$,  and  the superscript  $T$
indicates matrix transposition.  Suppose  that the vector field \( \mathcal{D}
\) can be decomposed as
\begin{equation}
\label{dec}
\mathcal{D}\left( \varrho \right) =-\frac{1}{2}K\nabla_\varrho V
\left( \varrho \right)
+\mathcal{A}\left( \varrho \right) ,\qquad \textrm{with } ~ \left\langle
K\nabla_\varrho V\, ,\, \mathcal{A}\right\rangle = 0 ~,
\label{dec_dissipative}
\end{equation}
and   let   \(  \hat{\varrho}   \)   be   a  minimum   of   \(   V  \),   with
$V(\hat{\varrho})=0$.  This  decomposition separates dissipative contributions
to  $\mathcal{D}$ from non-dissipative  ones, and  is considered  in diffusion
processes  described  by  finite  dimensional Langevin  equations  \cite{FW}.  
Integrating by parts, the ``entropy'' functional can be written as
\begin{eqnarray}
\hskip -50pt
I_{\rm path} \left( \varrho \right) 
=\frac{1}{2}\int _{t_{i}}^{t_{f}}\left\langle \dot{\varrho }+\frac{1}{2}K
\nabla_\varrho V-\mathcal{A}\, ,\, \dot{\varrho }+\frac{1}{2}K
\nabla_\varrho V-\mathcal{A}\right\rangle \, \mathrm{d}t \label{firstl}
\\
\hspace {-12pt}= \frac{1}{2}\int _{t_{i}}^{t_{f}}\left\langle \dot{\varrho }-
\frac{1}{2}K\nabla_\varrho V-\mathcal{A}\, ,\, \dot{\varrho }-\frac{1}{2}
K\nabla_\varrho V-\mathcal{A}\right\rangle \, \mathrm{d}t+
\left[ V\left( \varrho _{f}\right) -V\left(
\varrho _{i}\right) \right] \, 
\label{jt1t2}
\end{eqnarray}
whose last term  has no variation. Hence, two kinds  of evolution are possible
for  $\varrho$:  the relaxations  converging  to  \(  \hat{\varrho} \),  which
minimize (\ref{firstl}) and obey (\ref{xdotb}), and the fluctuations away from
\( \hat{\varrho} \), which minimize (\ref{jt1t2}), i.e.\ 
\begin{equation}
\label{invhydr}
\dot{\varrho} = - \mathcal{D}^{*}\left( \varrho \right) =
\frac{1}{2}K\nabla_\varrho V+\mathcal{A}\left( \varrho \right) =  -\mathcal{D}
+ 2 \mathcal{A} ~.
\end{equation}
The  qualitative properties  of the  deterministic dynamics  do not  depend on
$\mathcal{A}$, as  the time derivative of  the Lyapunov function  $V$ does not
depend on $\mathcal{A}$:
\begin{equation}
\dot{V}(\varrho) =\nabla_\varrho V \left( \varrho \right)
\cdot\mathcal{D}\left( \varrho \right)-\frac 1 2\langle K \nabla_\varrho V \left( \varrho \right),K \nabla_\varrho V \left( \varrho \right)\rangle\le 0 ~.
\end{equation}
Thus,  taking  \(  \varrho_{i}=\hat{\varrho}  \), \(  \varrho_{f}=\rho  \)  in
(\ref{jt1t2}), one  finds that  \( \mathcal{A} \)  does not contribute  to the
``entropy''  production,  while  the  asymmetry  between  normal  and  adjoint
dynamics,  which  implies the  macroscopic  irreversibility,  depends on  this
non-dissipative term.  It turns out that one peculiarity of the nonequilibrium
Lorentz gas  of Ref.\cite{GR04} is that  $\mathcal{A}$ tends to  zero when the
number of particles grows, which  explains the absence of temporal asymmetries
in Ref.\cite{GR04}. This is due to  the fact that the particles don't interact
with each other  and that the dynamics is  chaotic \cite{GRV-jonaa,GRV-jonab}. 
Because that  is a  rather special situation,  temporal asymmetries  should be
common in nonequilibrium systems \cite{GRV-jonaa,GRV-jonab,PSR}.

Thus, the separation of a  reversible part from an irreversible process, which
was  part of the  ``pseudo-thermostatic'' theories,  like Thomson's  theory of
thermoelectricity,  and was  considered rather  artificial  in the  past (cf.  
\cite{DeGroot} pp.3,4), results particularly revealing in the present context.

Note that,  because there  is no ``natural''  concept of FRP  in deterministic
dynamics,    several    notions    of    FRPs   have    been    proposed    in
Refs.\cite{GR04,GRV-jonaa,GRV-jonab,PSR}.

\section{Numerical and experimental tests}
\label{sec:tests}

The FR  for nonequilibrium  systems was proposed  and numerically  verified by
Evans, Cohen and Morris  in Ref.\cite{ECM}, where $\Omega=\Lambda$. After this
seminal paper, several  numerical tests have been devoted  to the $\zL$-FR, in
order  to  understand   how  properly  do  the  Chaotic   Hypothesis  and  the
$\Lambda$-FR describe models of physical  systems, as, in general, they do not
enjoy                   the                   Anosov                  property
\cite{BGG,GRS,LRB,BR01,SE2000,RM03,romans,DK,GZG,TG}.    Most  tests  concern,
instead, the $\zW$-FR, or functions  related to the dissipated energy, both in
transient        and       steady        states,       {\it        e.g.}       
Refs.\cite{BGG,LLP1,RS,GRS,LRB,BR01,earlierpapersA,earlierpapersB,generalized,SE2000,HeatFlow,chatelier}. 
{Let us consider some of these works.}

Lepri, Livi and Politi considered the nonequilibrium FPU chain \cite{LLP1}, in
which the first and last  oscillators are coupled to Nos\'e-Hoover thermostats
at different temperatures  $T_L$ and $T_R$, and the energy  fluxes at the left
and right ends of the chain, $j_L$ and $j_R$, are given by
\begin{equation} \label{eq:llp-1}
j_{L,R} = -\langle\zeta_{L,R}\rangle T_{L,R} \ ,
\end{equation}
where  $\langle\zeta_{L,R}\rangle$  is the  mean  value  of the  corresponding
effective  momentum of  the  thermostat. Since$j_L=-j_R=j$  in the  stationary
state, Eq.(\ref{eq:llp-1}) yields
\begin{equation} \label{eq:llp-2}
\langle\zeta_L\rangle + \langle\zeta_R\rangle = j\left(\frac{1}{T_R} -
  \frac{1}{T_L}\right) 
\end{equation}
which is equivalent to the  global entropy production, as obtained from linear
response.    For  the   Nos\'e-Hoover   thermostat  $\langle\zeta_L\rangle   +
\langle\zeta_R\rangle$  equals $\langle  \zL \rangle$,  but  the instantaneous
values and the PDF of $\zL$ and $j$ are not equal \cite{MMR1}.  Interestingly,
the energy flux was found to obey the FR.

{In Ref.\cite{LRB}, the  nonequilibrium  Ehrenfest
  wind-tree model is studied. Despite the  lack of chaos, the wind-tree model,
  with  small external fields}  and isokinetic  Gaussian thermostat,  has long
quasi-steady  transients, in which  the dynamics  looks random  and a  sort of
steady state \WFR holds, although the asymptotic state is a periodic orbit.  A
similar result is found in the polygonal billiards of Ref.\cite{BR01}.

In Ref.\cite{RM03}, the modified $\Lambda$-FR, with $c$ given by Eq.(\ref{pairs}), was found
consistent   with    highly   dissipative    SLLOD   systems   but,    as   in
Refs.\cite{BGG,GZG},  a direct  test could  not be  performed, because  of the
scarce statistics of negative fluctuations.  However, assuming the validity of
a simple scaling for the PDFs of the fluctuations, an indirect verification of
Eq.(\ref{pairs}) was obtained. A direct verification of a modified $\zL$-FR is
found in Ref.\cite{GRS},  for the GNS models of  Eq.(\ref{GNS}), with very few
modes but large Reynolds number.  The  only particle system in which an excess
of  negative Lyapunov  exponents  has been  obtained  without suppressing  the
negative  fluctuations   of  $\zL$   is  the  low   dimensional  Nos\'e-Hoover
thermostated oscillator of Ref.\cite{stephennew},  for which the standard \LFR
was verified.

The extended  FR of  Van Zon and  Cohen \cite{VZCa,VZCb} has  been numerically
verified for  overdamped Langevin particles and other  stochastic models, such
as the Markov chain and the  granular fluids of \cite{PRV}. A verification for
other   confining  potentials   and  trap   motions  has   been   obtained  in
Ref.\cite{Maesrecent}.   Gilbert has  studied the  Nos\'e-Hoover thermostatted
Lorentz  gas  \cite{TG}, previously  considered  in  \cite{DK},  in which  one
particle  diffuses in  a  billiard and  is  subjected to  an external  uniform
electric field and  a Nos\'e-Hoover thermostat. Differently from  the IE case,
the  Nos\'e-Hoover  model  has   unbounded  $\zL$.   Gilbert  found  that  the
fluctuations of $\zL$ follow (up  to finite size effects) Eq.(\ref{eq:VZCFT}). 
{However,  it is  not  clear which  class  of systems  obeys the  modified
  $\zL$-FR,  since models  with Nos\'e-Hoover  thermostats, such  as
  those of \cite{stephennew,MMR1}, obey the standard FR.}

Among  the numerical  studies  on temporal  asymmetries  of FRPs,  the one  of
Ref.\cite{Denisov} concerns  the fluctuations of  cross correlation functions,
not considered in Refs.\cite{GR04,GRV-jonaa,GRV-jonab,PSR}.

Experimental tests of FRs pose  difficult problems.  The first such experiment
was presented by Ciliberto and  Laroche, in 1998 \cite{ciliberto-1}, where the
temperature  fluctuations in a  fluid undergoing  Rayleigh-B\'enard convection
were found to obey a linear law similar to the FR, but with a different slope.
A  delicate  point of  Ref.\cite{ciliberto-1},  is  that  the fluctuations  in
temperature  were  considered  proportional  to the  fluctuations  in  entropy
production.

In 2002  Wang et al.\ experimentally  studied the
fluctuations of work done on a  colloidal particle dragged through water by an
optical trap, and verified an  integrated form of the transient $\zW$-FR. This
experiment motivated  Ref.\cite{VZCa,VZCb}.  An experimental  verification of
the steady state $\zW$-FR followed \cite{CRWSSE}.

In 2004 Feitosa and Menon studied a mechanically driven inelastic granular gas
in  a  fluidized  steady  state  \cite{feitosa}.  They  considered  the  power
fluctuations  in  a  subvolume  of  the  box  containing  the  granular  gas.  
Identifying the entropy  production as the quotient between  the power and the
effective temperature, they verified a local version of $\zW$-FR.

Garnier and  Ciliberto, in  2005, studied the  fluctuations of  the dissipated
power of  an electric dipole, consisting  of a resistor  connected in parallel
with  a capacitor,  and driven  out of  equilibrium by  an  electrical current
through the  circuit \cite{exptssa,exptssb}.  The work  and heat fluctuations,
both related to the fluctuations of the power dissipated by the resistor, were
considered: the work fluctuations  satisfy the $\Omega$-FR with high accuracy,
while  the PDF  of heat  has non-Gaussian  tails, and  is consistent  with the
extended FR of Van Zon and Cohen.

Shang  et al.\  studied  the fluctuations  of  a local  entropy production  in
turbulent thermal convection \cite{shang}.  {They considered a cylindrical
  cell filled with  water.} The steady state $\zW$-FR  was confirmed measuring
the velocity and temperature fields.

Tietz et al.\ measured the entropy production for a single two-level system, a
defect center in natural  IIa-type diamond \cite{tietz06}.  Using fluorescence
spectroscopy,  they studied  the transitions  between ``dark''  and ``bright''
states and, following \cite{seifert05}, showed that their ``stochastic entropy
production'' satisfies a kind of FR.

In  2006,  Douarche  et al.\  studied  the  steady  state and  transient  work
fluctuations of a  damped harmonic oscillator that is  kept out of equilibrium
by an external force \cite{ciliberto-2}. They considered a torsion pendulum in
a cell  filled with a solution  of water-glycerol, and  measured optically its
torsional motion.  A time dependent external torque was applied and controlled
by  an electric  current,  and the  fluctuations  of work  were studied.   The
transient FR was confirmed for any averaging time. The steady-state version is
observed  to converge,  although  in a  complex  fashion that  depends on  the
external driving.   Douarche et  al.\ had previously  experimentally confirmed
the  Jarzynski   and  Crooks  equalities   in  the  same   experimental  setup
\cite{ciliberto-3}.

Blickle et al.\ tested the validity of the Jarzynski and Crooks equalities for
a colloidal particle in a time-dependent nonharmonic potential \cite{blickle}.
Their  experiment   consisted  in  an  aqueous   suspension  of  micrometrical
polystyrene  beads. Using  optical tweezers  they drove  one of  the colloidal
particles between  two equilibrium states and  found that the  work exerted on
the particle verifies the Jarzynski and Crooks relations.

The Jarzynski  equality has found  applicability in molecular  and biophysical
experiments, because  it can be used  to estimate the  equilibrium free energy
out of measurements  of dissipated work, in nonequilibrium  processes. This is
particularly useful in systems for which  no other method to estimate the free
energy exists. The Jarzynski Equality  was first confirmed in 2002 by Liphardt
et al.\ in measurements of the dissipated work in folding-unfolding process of
a  single molecule  of  RNA \cite{liphardt}.   Collin  et al.\  experimentally
confirmed the  Crooks relation near and far  from equilibrium \cite{ritort-1},
using optical tweezers and measuring the dissipative work during the unfolding
and refolding of a small RNA molecule.

A  relation which  is often  mentioned in  connection with  the  Jarzynski and
Crooks relations is the steady-state  equality of Hatano and Sasa \cite{HS01},
which  has  been verified  in  \cite{trepagnier}.   For  more on  experimental
verifications of FRs, see Refs.\cite{ritort,bustamante,PGaspnanotech}.

\section{Concluding remarks}
\label{sec:conclusions}

A  unifying picture  of  nonequilibrium  physics is  emerging,  thanks to  the
development of theories describing the nonequilibrium fluctuations, whose role
appears to  be at least  as fundamental as  that of equilibrium  fluctuations. 
There  are two kinds  of FRs:  transient and  steady state  FRs, which  are of
totally different nature.  Besides  being interesting for conceptual reasons,
both kinds  of relations  are important in  the description of  small systems,
such as nanotechnological devices and biological systems.

The transient FRs connect in a striking fashion equilibrium and nonequilibrium
properties of physical systems, in  that they consider at once the statistical
properties   of  equilibrium  states   and  nonequilibrium   dynamics.   Their
predictions  describe the statistics  of ensembles  of experiments,  are valid
under extremely  wide conditions, and  can be verified  by a large  variety of
physical systems.  The steady state  relations, on the other hand, concern the
asymptotic  statistics generated  by a  single system  evolution, if  a steady
state  is  reached.

Despite the available rigorous derivations of such relations require 
very restrictive conditions, which are hardly met by any 
  system  of  physical interest, the steady state FRs appear to hold for a wide
class of systems. Indeed, the  analysis  of  the physical  mechanisms
underlying the validity of these relations explains why they should be verified 
as widely as  common
thermodynamic relations can. As Section 5 shows, the steady state $\zW$-FR and
its consequences can be obtained only from time reversibility and from the
$\zW$-autocorrelation decay. With these ingredients, indeed, the $\zW$-FR is rigorously 
estabilished \cite{ESR2}.  The above  analysis cannot contradict the theory
based   on   axiom   C   systems,   because  it   rests   on   exact   results
--Eq.(\ref{SSESFT})  in  particular-- but  it  raises  various questions.  For
instance, the fact that the decay  of correlations of axiom C systems does not
imply the  $\zW$-autocorrelation decay required  by the steady  state $\zW$-FR
deserves  further investigation. Also,  in the  linear regime  of Irreversible
Thermodynamics, the $\zW$-autocorrelation decay  is required for the transport
coefficients to exist, but the  meaning of (\ref{aveBDD}) far from equilibrium
has still to be fully understood, although it is consistent with the available
data.  The analysis based on the decay of the $\zW$-autocorrelation leads also
to  a number  of  predictions \cite{ESR2},  most  of which  have  still to  be
considered in experimental tests, and  attributes a lower importance to strong
chaos  than to  correlations  decay of  small  sets of  observables, which  is
favoured by the large number of degrees of freedom \cite{GRV-jonab,Kantz}.
 
The interplay of deterministic and stochastic approaches is also quite useful.
For instance,  we have  seen how the  result of  Evans, Cohen and  Morriss for
deterministic  systems has  motivated much  research on  stochastic processes,
while the  work by  Bertini, De Sole,  Gabrielli, Jona-Lasinio and  Landim has
motivated  the  study  of  otherwise unexpected  properties  of  deterministic
systems.

Having understood the physical mechanisms  underlying the validity of the FRs,
the present theory may be further developed in directions which aim to clarify
various open questions, which are  both of mathematical and physical interest. 
Among those  mentioned at the end  of Sections 1, 4  and 5, let  us recall the
decay of correlations with respect to  the equilibrium and to the steady state
measures,  the physical  relevance of  axiom  C systems,  the construction  of
dynamical systems which rigorously  obey (\ref{aveBDD}), and the properties of
different thermostatting mechanisms.  The interaction between mathematical and
physical approaches seems particularly necessary to explore these new lands.

\section*{Acknowledgment}
We are grateful to A Vulpiani for insightful remarks, and thank Fondazione CRT
for financial support.

\section*{References}
\bibliography{FlucThms}{}

\begin{thebibliography}{100}

\bibitem{AE05}
A~Einstein.
\newblock The motion of elements suspended in static liquids as claimed in the
  molecular kinetic theory of heat.
\newblock {\em Ann. Physik}, 17:549, 1905.

\bibitem{AE10}
A~Einstein.
\newblock Theory of opalescence of homogeneous liquids and mixtures of liquids
  in the vicinity of the critical state.
\newblock {\em Ann. Physik}, 33:1275, 1910.

\bibitem{LSO27}
L~S Ornstein.
\newblock On the theory of {B}rownian motion for systems out of equilibrium.
\newblock {\em Z. Phys.}, 41:848, 1927.

\bibitem{HN28}
H~Nyquist.
\newblock Thermal agitation of electric charge in conductors.
\newblock {\em Phys. Rev.}, 32:110, 1928.

\bibitem{LO31a}
L~Onsager.
\newblock Reciprocal relations in irreversible processes. {I}.
\newblock {\em Phys. Rev.}, 37:405, 1931.

\bibitem{LO31b}
L~Onsager.
\newblock Reciprocal relations in irreversible processes. {II}.
\newblock {\em Phys. Rev.}, 38:2265, 1931.

\bibitem{CWG51}
H~B Callen and T~A Welton.
\newblock Irreversibility and generalized noise.
\newblock {\em Phys. Rev.}, 83:34, 1951.

\bibitem{CWG52}
H~B Callen and R~F Greene.
\newblock On a theorem of irreversible thermodynamics.
\newblock {\em Phys. Rev.}, 86:702, 1952.

\bibitem{MSG51}
M~S Green.
\newblock Brownian motion in a gas of noninteracting molecules.
\newblock {\em J. Chem. Phys.}, 19:1036, 1951.

\bibitem{MSG52}
M~S Green.
\newblock Markoff random processes and the statistical mechanics of
  time-dependent phenomena.
\newblock {\em J. Chem. Phys.}, 20:1281, 1952.

\bibitem{MSG54}
M~S Green.
\newblock Markoff random processes and the statistical mechanics of
  time-dependent phenomena. {II}. {I}rreversible processes in fluids.
\newblock {\em J. Chem. Phys.}, 22:398, 1954.

\bibitem{RK57}
R~Kubo.
\newblock Statistical-mechanical theory of irreversible processes. {I}. general
  theory and simple applications to magnetic and conduction problems.
\newblock {\em J. Phys. Soc. Jap.}, 12:570, 1957.

\bibitem{OM53a}
L~Onsager and S~Machlup.
\newblock Fluctuations and irreversible processes.
\newblock {\em Phys. Rev.}, 91:1505, 1953.

\bibitem{OM53b}
S~Machlup and L~Onsager.
\newblock Fluctuations and irreversible process. {II. S}ystems with kinetic
  energy.
\newblock {\em Phys. Rev.}, 91:1512, 1953.

\bibitem{AW67}
B~J Alder and T~E Wainwright.
\newblock Velocity autocorrelations for hard spheres.
\newblock {\em Phys. Rev. Lett.}, 18:988, 1967.

\bibitem{KS68}
L~P Kadanoff and J~Swift.
\newblock Transport coefficients near the liquid-gas critical point.
\newblock {\em Phys. Rev.}, 166:89, 1968.

\bibitem{LRCa}
I~Procaccia, D~Ronis, and I~Oppenheim.
\newblock Light scattering from nonequilibrium stationary states: {t}he
  implication of broken time-reversal symmetry.
\newblock {\em Phys. Rev. Lett.}, 42:287, 1979.

\bibitem{LRCb}
T~R Kirkpatrick, E~G~D Cohen, and J~R Dorfman.
\newblock Kinetic theory of light scattering from a fluid not in equilibrium.
\newblock {\em Phys. Rev. Lett.}, 42:862, 1979.

\bibitem{Spohnbook}
H~Spohn.
\newblock {\em {Large Scale Dynamics of Interacting Particles}}.
\newblock Text, monographs in Physics (Heidelberg: Springer-Verlag), 1991.

\bibitem{HT75}
P~H\"anggi and H~Thomas.
\newblock Linear response and fluctuation theorems for nonstationary
  stochastic-processes.
\newblock {\em Z. Physik B}, 22:295, 1975.

\bibitem{HT78}
P~H\"anggi.
\newblock Stochastic-processes .2. response theory and fluctuation theorems.
\newblock {\em Helvetica Phys. Acta}, 51:202, 1978.

\bibitem{Viss}
W~M Visscher.
\newblock Transport processes in solids and linear-response theory.
\newblock {\em Phys. Rev. A}, 10:2461, 1974.

\bibitem{DuLi}
J~W Dufty and M~J Lindenfeld.
\newblock Non-linear transport in the {B}oltzmann limit.
\newblock {\em J. Stat. Phys.}, 20:259, 1979.

\bibitem{Coh}
E~G~D Cohen.
\newblock Kinetic-theory of non-equilibrium fluids.
\newblock {\em Physica A}, 118:17, 1983.

\bibitem{EMTTCF}
G~P Morriss and D~J Evans.
\newblock Application of transient correlation functions to shear flow far from
  equilibrium.
\newblock {\em Phys. Rev. A}, 35:792, 1987.

\bibitem{FIVa}
M~Falcioni, S~Isola, and A~Vulpiani.
\newblock Correlation functions, relaxation properties in chaotic dynamics,
  statistical mechanics.
\newblock {\em Phys. Lett. A}, 144:341, 1990.

\bibitem{FIVb}
G~Boffetta, G~Lacorata, S~Musacchio, and A~Vulpiani.
\newblock Relaxation of finite perturbations: Beyond the fluctuation-response
  relation.
\newblock {\em Chaos}, 13:806, 2003.

\bibitem{SRBdiff}
D~Ruelle.
\newblock Differentiation of {SRB} states.
\newblock {\em Comm. Math. Phys.}, 187:227, 1997.

\bibitem{ECM}
D~J Evans, E~G~D Cohen, and G~P Morriss.
\newblock Probability of second law violations in shearing steady flows.
\newblock {\em Phys. Rev. Lett.}, 71:2401, 1993.

\bibitem{earlierpapersA}
D~J Evans and D~J Searles.
\newblock Equilibrium microstates which generate second law violating steady
  states.
\newblock {\em Phys. Rev. E}, 50:1645, 1994.

\bibitem{earlierpapersB}
D~J Evans and D~J Searles.
\newblock Steady sates, invariant measures, response theory.
\newblock {\em Phys. Rev. E}, 52:5839, 1995.

\bibitem{GCa}
G~Gallavotti and E~G~D Cohen.
\newblock Dynamical ensembles in nonequilibrium statistical mechanics.
\newblock {\em Phys. Rev. Lett.}, 94:2694, 1995.

\bibitem{GCb}
G~Gallavotti and E~G~D Cohen.
\newblock Dynamical ensembles in stationary states.
\newblock {\em J. Stat. Phys.}, 80:931, 1995.

\bibitem{GG96}
G~Gallavotti.
\newblock Extension of {O}nsager's reciprocity to large fields, the chaotic
  hypothesis.
\newblock {\em Phys. Rev. Lett.}, 77:4334, 1996.

\bibitem{GR97}
G~Gallavotti and D~Ruelle.
\newblock {SRB} states and nonequilibrium statistical mechanics close to
  equilibrium.
\newblock {\em Comm. Math. Phys.}, 190:279, 1997.

\bibitem{ESR}
D~J Evans, D~J Searles, and L~Rondoni.
\newblock On the application of the {Gallavotti-Cohen} fluctuation relation to
  thermostatted steady states near equilibrium.
\newblock {\em Phys. Rev. E}, 71:056120, 2005.

\bibitem{Kurchanlast}
J~Kurchan.
\newblock Nonequilibrium work relations.
\newblock 2005.
\newblock http://arXiv.org/cond-mat/0511073.

\bibitem{JB02}
J~Bellisard.
\newblock Coherent, dissipative transport in aperiodic solids: an overview.
\newblock 2002.
\newblock in Dynamics of dissipation, Lecture Notes in Physics 597
  P.Garbaczewski, R.Olkiewicz Eds., (Springer Verlag, Berlin.).

\bibitem{BuniLP}
S~Lansel, M~A Porter, and L~A Bunimovich.
\newblock One-particle and few-particle billiards.
\newblock {\em CHAOS}, 16:013129, 2006.

\bibitem{JR06}
O~G Jepps and L~Rondoni.
\newblock Thermodynamics and complexity of simple transport phenomena.
\newblock {\em J. Phys. A}, 39:1311, 2006.

\bibitem{MH}
B~Moran and W~G Hoover.
\newblock Diffusion in a periodic {L}orentz gas.
\newblock {\em J. Stat. Phys.}, 48:709, 1987.

\bibitem{CELS}
N~I Chernov, G~L Eyink J~L Lebowitz, and Ya~G Sinai.
\newblock Steady-state electrical conduction in the periodic {L}orentz gas.
\newblock {\em Comm. Math. Phys.}, 154:569, 1993.

\bibitem{LNRM}
J~Lloyd, M~Niemeyer, L~Rondoni, and G~P Morriss.
\newblock The nonequilibrium {L}orentz gas.
\newblock {\em Chaos}, 5:536, 1995.

\bibitem{BSC}
L~A Bunimovich, Ya~G Sinai, and N~I Chernov.
\newblock Statistical properties of two-dimensional hyperbolic billiards.
\newblock {\em Russian Math. Surveys}, 46:47, 1991.

\bibitem{bustamante}
C~Bustamante, J~Liphardt, and F~Ritort.
\newblock The nonequilibrium thermodynamics of small systems.
\newblock {\em Physics Today}, 58:43, 2005.

\bibitem{ESR2}
D~J Evans, D~J Searles, and L~Rondoni.
\newblock The steady state fluctuation relation for the dissipation function.
\newblock 2007.
\newblock submitted.

\bibitem{BGG}
F~Bonetto, G~Gallavotti, and P~L Garrido.
\newblock Chaotic principle: {a}n experimental test.
\newblock {\em Physica D}, 105:226, 1997.

\bibitem{axiomC}
F~Bonetto and G~Gallavotti.
\newblock Reversibility, coarse graining, the chaoticity principle.
\newblock {\em Comm. Math. Phys.}, 189:263, 1997.

\bibitem{GRS}
G~Gallavotti, L~Rondoni, and E~Segre.
\newblock {Lyapunov spectra, nonequilibrium ensembles equivalence in 2D fluid
  mechanics}.
\newblock {\em Physica D}, 187:338, 2004.

\bibitem{stephennew}
S~R Williams, D~J Searles, and D~J Evans.
\newblock Numerical study of the steady state fluctuation relations far from
  equilibrium.
\newblock {\em J. Chem. Phys.}, 124:194102, 2006.

\bibitem{review}
D~J Evans and D~J Searles.
\newblock The fluctuation theorem.
\newblock {\em Adv. Phys.}, 52:1529, 2002.

\bibitem{Khinchin}
A~I Khinchin.
\newblock {\em {Mathematical Foundations of Statistical Mechanics}}.
\newblock Dover Publications, New York, 1949.

\bibitem{EM}
D~J Evans and G~P Morriss.
\newblock {\em {Statistical Mechanics of Nonequlibrium Liquids}}.
\newblock New York: Academic Press, 1990.

\bibitem{POEinECM2a}
W~Parry.
\newblock Synchronisation of canonical measures for hyperbolic attractors.
\newblock {\em Comm. Math. Phys.}, 106:267, 1986.

\bibitem{POEinECM2b}
W~N Vance.
\newblock Unstable periodic orbits, transport properties of nonequilibrium
  steady states.
\newblock {\em Phys. Rev. Lett.}, 69:1356, 1992.

\bibitem{generalized}
D~J Searles, G~Ayton, and D~J Evans.
\newblock Generalised fluctuation formula.
\newblock {\em AIP Conference Series}, 519:271, 2000.

\bibitem{SE2000}
D~J Searles and D~J Evans.
\newblock Ensemble dependence of the transient fluctuation theorem.
\newblock {\em J. Chem. Phys.}, 113:3503, 2000.

\bibitem{StephenPRE}
S~R Williams, D~J Searles, and D~J Evans.
\newblock Thermostat invariance of the transient fluctuation theorem.
\newblock {\em Phys. Rev. E}, 70:066113, 2004.

\bibitem{WSMSE}
G~M Wang, E~M Sevick, E~Mittag, D~J Searles, and D~J Evans.
\newblock Experimental demonstration of violations of the second law of
  thermodynamics for small systems and short time scales.
\newblock {\em Phys. Rev. Lett.}, 89:050601, 2002.

\bibitem{romans}
F~Zamponi, G~Ruocco, and L~Angelani.
\newblock Fluctuations of entropy production in the isokinetic ensemble.
\newblock {\em J. Stat. Phys.}, 115:1655, 2004.

\bibitem{GZG}
A~Giuliani, F~Zamponi, and G~Gallavotti.
\newblock Fluctuation relation beyond linear response theory.
\newblock {\em J. Stat. Phys.}, 119:909, 2005.

\bibitem{AustJChem}
D~J Searles and D~J Evans.
\newblock Fluctuation relations for nonequilibrium systems.
\newblock {\em Aust. J. Chem.}, 57:1119, 2004.

\bibitem{DK}
M~Dolowschiak and Z~Kovacs.
\newblock {Fluctuation formula in the Nos\'e-Hoover thermostated Lorentz gas}.
\newblock {\em Phys. Rev. E}, 71:025202, 2005.

\bibitem{exptssa}
G~M Wang, J~C Reid, D~M Carberry, D~R~M Williams, E~M Sevick, and D~J Evans.
\newblock Experimental study of the fluctuation theorem in a nonequilibrium
  steady state.
\newblock {\em Phys. Rev. E}, 71:046142, 2005.

\bibitem{exptssb}
N~Garnier and S~Ciliberto.
\newblock Nonequilibrium fluctuations in a resistor.
\newblock {\em Phys. Rev. E}, 71:060101, 2005.

\bibitem{LLP1}
S~Lepri, R~Livi, and A~Politi.
\newblock Energy transport in anharmonic lattices close to, far from
  equilibrium.
\newblock {\em Physica D}, 119:140, 1998.

\bibitem{TG}
T~Gilbert.
\newblock Fluctuation theorem applied to the {Nos\'e-Hoover} thermostated
  {L}orentz gas.
\newblock {\em Phys. Rev. E}, 73:035102, 2006.

\bibitem{GG-MPEJ}
G~Gallavotti.
\newblock Reversible {A}nosov diffeomorphisms, large deviations.
\newblock {\em Math. Phys. Electronic J.}, 1:1, 1995.

\bibitem{GGrevisited}
G~Gallavotti.
\newblock Fluctuation theorem revisited.
\newblock 2004.
\newblock http://arXiv.org/cond-mat/0402676.

\bibitem{EPRB}
J-P Eckmann, C-A Pillet, and L~Rey-Bellet.
\newblock Entropy production in nonlinear, thermally driven {H}amiltonian
  systems.
\newblock {\em J. Stat. Phys.}, 95:305, 1999.

\bibitem{RBT}
L~Rey-Bellet and L~E Thomas.
\newblock Fluctuations of the entropy production in anharmonic chains.
\newblock {\em Ann. H Poincar\'e}, 3:483, 2002.

\bibitem{TRVopen}
L~Rondoni, T~T\'el, and J~Vollmer.
\newblock Fluctuation theorems for entropy production in open systems.
\newblock {\em Phys. Rev. E}, 61:R4679, 2000.

\bibitem{AES}
G~Ayton, D~J Evans, and D~J Searles.
\newblock A local fluctuation theorem.
\newblock {\em J. Chem. Phys.}, 115:2033, 2001.

\bibitem{CM00}
C~Maes, F~Redig, and M~Verschuere.
\newblock From global to local fluctuation theorems.
\newblock {\em Moscow Math. J.}, 1:421, 2001.

\bibitem{Kurchan}
J~Kurchan.
\newblock Fluctuation theorem for stochastic dynamics.
\newblock {\em J. Phys. A}, 31:3719, 1998.

\bibitem{LS}
J~L Lebowitz and H~Spohn.
\newblock A {Gallavotti-Cohen}-type symmetry in the large deviation functional
  for stochastic dynamics.
\newblock {\em J. Stat. Phys.}, 95:333, 1999.

\bibitem{stochasticES}
D~J Searles and D~J Evans.
\newblock The fluctuation theorem for stochastic systems.
\newblock {\em Phys. Rev. E}, 60:159, 1999.

\bibitem{CM99}
C~Maes.
\newblock The fluctuation theorem as a {G}ibbs property.
\newblock {\em J. Stat. Phys.}, 95:367, 1999.

\bibitem{VZCa}
R~van Zon and E~G~D Cohen.
\newblock Extension of the fluctuation theorem.
\newblock {\em Phys. Rev. Lett.}, 91:110601, 2003.

\bibitem{VZCb}
R~van Zon and E~G~D Cohen.
\newblock Extended heat-fluctuation theorems for a system with deterministic
  and stochastic forces.
\newblock {\em Phys. Rev. E}, 69:056121, 2004.

\bibitem{BD}
T~Bodineau and B~Derrida.
\newblock Current fluctuations in nonequilibrium diffusive systems: an
  additivity principle.
\newblock {\em Phys. Rev. Lett.}, 92:180601, 2004.

\bibitem{BDSJLcurrent}
L~Bertini, A~De Sole, D~Gabrielli, G~Jona-Lasinio, and C~Landim.
\newblock Current fluctuations in stochastic lattice gases.
\newblock {\em Phys. Rev. Lett.}, 94:030601, 2005.

\bibitem{GAFLU}
G~Gallavotti.
\newblock Dynamical ensemble equivalence in fluid mechanics.
\newblock {\em Physica D}, 105:163, 1997.

\bibitem{RS}
L~Rondoni and E~Segre.
\newblock Fluctuations in two-dimensional reversibly damped turbulence.
\newblock {\em Nonlinearity}, 12:1471, 1999.

\bibitem{QKurchan}
J~Kurchan.
\newblock A quantum fluctuation theorem.
\newblock 2000.
\newblock http://arXiv.org/cond-mat/0007360.

\bibitem{MT}
T~Monnai.
\newblock Unified treatment of the quantum fluctuation theorem, the {J}arzynski
  equality in terms of microscopic reversibility.
\newblock {\em Phys. Rev. E}, 72:027102, 2005.

\bibitem{DRM}
A~De Roeck and C~Maes.
\newblock Steady state fluctuations of the dissipated heat for a quantum
  stochastic model.
\newblock 2004.
\newblock http://arXiv.org/cond-mat/0406004.

\bibitem{Mukamel}
M~Esposito, U~Harbola, and S~Mukamel.
\newblock Fluctuation theorem for counting-statistics in electron transport
  through quantum junctions.
\newblock 2007.
\newblock http://arXiv.org/cond-mat/0702376.

\bibitem{JW}
C~Jarzynski and D~K W\'ojcik.
\newblock Classical, quantum fluctuation theorems for heat exchange.
\newblock {\em Phys. Rev. Lett.}, 92:230602, 2004.

\bibitem{CJ}
C~Jarzynski.
\newblock Nonequilibrium equality for free energy differences.
\newblock {\em Phys. Rev. Lett.}, 78:2690, 1997.

\bibitem{GK}
G~E Crooks.
\newblock Path ensemble averages in systems driven far from equilibrium.
\newblock {\em Phys. Rev. E}, 61:2361, 2000.

\bibitem{HS01}
T~Hatano and S~Sasa.
\newblock Steady-state thermodynamics of {L}angevin systems.
\newblock {\em Phys. Rev. Lett.}, 86:3463, 2001.

\bibitem{PO98}
Y~Oono and M~Paniconi.
\newblock Steady state thermodynamics.
\newblock {\em Progr. Theor. Phys. Suppl.}, 130:29, 1998.

\bibitem{GGspringer}
G~Gallavotti.
\newblock {\em {Statistical Mechanics: a Short Treatise}}.
\newblock (Springer Verlag Berlin), 2000.

\bibitem{AT87}
M~P Allen and D~J Tildesley.
\newblock {\em {Computer Simulation of Liquids}}.
\newblock (New York: Oxford University Press), 1987.

\bibitem{WH91}
W~G Hoover.
\newblock {\em {Computational Statistical Mechanics}}.
\newblock Elsevier, 1991.

\bibitem{SEC98}
S~Sarman, D~J Evans, and P~T Cummings.
\newblock Recent developments in non-{N}ewtonian molecular dynamics.
\newblock {\em Physics Reports}, 305:1, 1998.

\bibitem{gauss}
K~F Gauss.
\newblock Nouveau principe de m\'ecanique.
\newblock {\em J. Reine Angewandte Math.}, 4:232, 1829.

\bibitem{lanczos}
C~Lanczos.
\newblock {\em {The Variational Principles of Mechanics}}.
\newblock Dover, New York, 1979.

\bibitem{NHa}
S~Nos\'e.
\newblock A unified formulation of the constant temperature molecular-dynamics
  methods.
\newblock {\em J. Chem. Phys}, 81:511, 1984.

\bibitem{NHb}
S~Nos\'e.
\newblock A molecular-dynamics method for the simulations in the canonical
  ensemble.
\newblock {\em Mol. Phys.}, 52:255, 1984.

\bibitem{NHc}
W~G Hoover.
\newblock Canonical dynamics: Equilibrium phase-space distributions.
\newblock {\em Phys. Rev. A}, 31:1695, 1985.

\bibitem{RUPhysToday}
D~Ruelle.
\newblock Conversations on nonequilibrium physics with an extraterrestrial.
\newblock {\em Physics Today}, 57:48, 2004.

\bibitem{RC98}
L~Rondoni and E~G~D Cohen.
\newblock Orbital measures in non-equilibrium statistical mechanics: the
  {O}nsager relations.
\newblock {\em Nonlinearity}, 11:1395, 1998.

\bibitem{LRladek}
L~Rondoni.
\newblock Deterministic thermostats, fluctuation relations.
\newblock 2002.
\newblock in Dynamics of dissipation, Lecture Notes in Physics 597
  P.Garbaczewski, R.Olkiewicz Eds., (Springer Verlag, Berlin.).

\bibitem{ESA}
D~J Evans and S~Sarman.
\newblock Equivalence of thermostatted nonlinear responses.
\newblock {\em Phys. Rev. E}, 48:65, 1993.

\bibitem{BTV}
T~Tel, J~Vollmer, and W~Breymann.
\newblock Transient chaos: The origin of transport in driven systems.
\newblock {\em Europhys. Lett.}, 35:659, 1996.

\bibitem{CR98}
E~G~D Cohen and L~Rondoni.
\newblock Note on phase space contraction and entropy production in
  thermostatted {H}amiltonian systems.
\newblock {\em Chaos}, 8:357, 1998.

\bibitem{BLnonequiv}
F~Bonetto and J~L Lebowitz.
\newblock Thermodynamic entropy production fluctuation in a two-dimensional
  shear flow model.
\newblock {\em Phys. Rev. E.}, 64:056129., 2001.

\bibitem{EMequiva}
D~J Evans and G~P Morriss.
\newblock Equilibrium time correlation-functions under {G}aussian isothermal
  dynamics.
\newblock {\em Chem. Phys.}, 87:451, 1984.

\bibitem{EMequivb}
D~J Evans and G~P Morriss.
\newblock Non-{N}ewtonian molecular dynamics.
\newblock {\em Comput. Phys. Rep.}, 1:297, 1984.

\bibitem{LBC}
S~Y Liem, D~Brown, and J~H~R Clarke.
\newblock Investigation of the homogeneous-shear
  nonequilibrium-molecular-dynamics method.
\newblock {\em Phys. Rev. A}, 45:3706., 1992.

\bibitem{HAHDG}
Wm~G Hoover, K~Aoki, C~G Hoover, and S~V~De Groot.
\newblock Time-reversible deterministic thermostats.
\newblock {\em Physica D}, 187:253, 2004.

\bibitem{HP04}
H~A Posch and Wm~G Hoover.
\newblock Large-system phase-space dimensionality loss in stationary heat
  flows.
\newblock {\em Physica D}, 187:281, 2004.

\bibitem{AK}
K~Aoki and D~Kusnezov.
\newblock {Lyapunov exponents, the extensivity of dimensional loss for systems
  in thermal gradients}.
\newblock {\em Phys. Rev. E}, 68:056204, 2003.

\bibitem{MP99}
F~M\"uller-Plathe.
\newblock Reversing the perturbation in nonequilibrium molecular dynamics: An
  easy way to calculate the shear viscosity of fluids.
\newblock {\em Phys. Rev. E}, 59:4894, 1999.

\bibitem{DRthermolim}
D~Ruelle.
\newblock A remark on the equivalence of isokinetic and isoenergetic
  thermostats in the thermodynamic limit.
\newblock {\em J. Stat. Phys.}, 100:757, 2000.

\bibitem{Gentile}
G~Gentile.
\newblock Large deviations for {A}nosov flows.
\newblock {\em Forum Math.}, 10:89, 1998.

\bibitem{sinaibook}
Ya~G Sinai.
\newblock {\em {Lectures in Ergodic Theory}}.
\newblock Lecture Notes in Mathematics, (Princeton University Press), 1977.

\bibitem{ECSB}
D~J Evans, E~G~D Cohen, D~J Searles, and F~Bonetto.
\newblock Note on the {Kaplan-Yorke} dimension and linear transport
  coefficients.
\newblock {\em J. Stat. Phys.}, 101:17, 2000.

\bibitem{DRdiff}
D~Ruelle.
\newblock Smooth dynamics, new theoretical ideas in nonequilibrium statistical
  mechanics.
\newblock {\em J. Stat. Phys.}, 95:393, 1999.

\bibitem{GGCHAOS}
G~Gallavotti.
\newblock Chaotic dynamics, fluctuations, nonequilibrium ensembles.
\newblock {\em Chaos}, 8:384, 1998.

\bibitem{ECM1}
D~J Evans, E~G~D Cohen, and G~P Morriss.
\newblock Viscosity of a simple fluid from its maximal {L}yapunov exponents.
\newblock {\em Phys. Rev. A}, 42:5990, 1990.

\bibitem{DMpairinga}
C~P Dettmann and G~P Morriss.
\newblock Proof of {L}yapunov exponent pairing for systems at constant kinetic
  energy.
\newblock {\em Phys. Rev. E}, 53:R5545, 1996.

\bibitem{DMpairingb}
C~P Dettmann and G~P Morriss.
\newblock {Hamiltonian formulation of the Gaussian isokinetic thermostat}.
\newblock {\em Phys. Rev. E}, 54:2495, 1996.

\bibitem{RM03}
L~Rondoni and G~P Morriss.
\newblock Large, fluctuations and axiom-{C} structures in deterministically
  thermostatted systems.
\newblock {\em Open Syst. Information Dynam.}, 10:105, 2003.

\bibitem{GGlocal}
G~Gallavotti.
\newblock A local fluctuation theorem.
\newblock {\em J. Phys. A}, 263:39, 1999.

\bibitem{GG-OM1}
G~Gallavotti.
\newblock Fluctuation patterns and conditional reversibility in nonequilibrium
  system.
\newblock {\em Ann. Inst. H. Poincar\'e}, 70:429, 1999.

\bibitem{GG-OM2}
G~Gallavotti.
\newblock Large deviations, fluctuation theorem, {Onsager-Machlup} theory in
  nonequilibrium statistical mechanics.
\newblock 2002.
\newblock http://ipparco.roma1.infn.it/.

\bibitem{LRB}
S~Lepri, L~Rondoni, and G~Benettin.
\newblock The {Gallavotti-Cohen} fluctuation theorem for a nonchaotic model.
\newblock {\em J. Stat. Phys.}, 99:857, 2000.

\bibitem{BR01}
G~Benettin and L~Rondoni.
\newblock A new model for the transport of particles in a thermostatted system.
\newblock {\em Math. Phys. Electronic J.}, 7:3, 2001.

\bibitem{DJE03}
D~J Evans.
\newblock A non-equilibrium free energy theorem for deterministic systems.
\newblock {\em Molecular Phys.}, 101:1551, 2003.

\bibitem{vzc}
R~van Zon and E~G~D Cohen.
\newblock Stationary and transient work-fluctuation theorems for a dragged
  {B}rownian particle.
\newblock {\em Phys. Rev. E}, 67:046102, 2003.

\bibitem{BGGZ}
F~Bonetto, G~Gallavotti, A~Giuliani, and F~Zamponi.
\newblock Chaotic hypothesis, fluctuation theorem, singularities.
\newblock {\em J. Stat. Phys.}, 123:39, 2006.

\bibitem{MMR1}
C~Mejia-Monasterio and L~Rondoni.
\newblock On the fluctuation theorem for a locally nose-hoover thermostated
  systems.
\newblock 2007.
\newblock (in preparation).

\bibitem{Farago}
J~Farago.
\newblock Injected power fluctuations in {L}angevin equation.
\newblock {\em J. Stat. Phys.}, 107:781, 2002.

\bibitem{Maesrecent}
M~Baiesi, T~Jacobs, C~Maes, and N~S Skantzos.
\newblock Fluctuation symmetries for work, heat.
\newblock {\em Phys. Rev. E}, 74:021111, 2006.

\bibitem{PRV}
A~Puglisi, L~Rondoni, and A~Vulpiani.
\newblock Relevance of initial and final conditions for the fluctuation
  relation in {M}arkov processes.
\newblock {\em J. Stat. Mech.}, page P08010, 2006.
\newblock http://arXiv.org/cond-mat/0606526.

\bibitem{germans}
R~J Harris, A~R\'akos, and M~Sch\"utz.
\newblock Breakdown of {Gallavotti-Cohen} symmetry for stochastic dynamics.
\newblock {\em Europhys. Lett.}, 75:227, 2006.

\bibitem{visco}
P~Visco.
\newblock Work fluctuations for a {B}rownian particle between two thermostats.
\newblock {\em J. Stat. Mech.}, page P06006, 2006.

\bibitem{CRWSSE}
D~M Carberry, J~C Reid, G~M Wang, E~M Sevick, D~J Searles, and D~J Evans.
\newblock Fluctuations and irreversibility: an experimental demonstration of a
  second-law-like theorem using a colloidal particle held in an optical trap.
\newblock {\em Phys. Rev. Lett.}, 92:140601, 2004.

\bibitem{ESR3}
D~J Evans, D~J Searles, and G~P Morriss.
\newblock (in preparation).

\bibitem{ESdissir}
D~J Evans and D~J Searles.
\newblock The dissipation theorem.
\newblock 2007.
\newblock (submitted).

\bibitem{CG99}
E~G~D Cohen and G~Gallavotti.
\newblock Note on two theorems in nonequilibrium statistical mechanics.
\newblock {\em J. Stat. Phys.}, 96:1343, 1999.

\bibitem{CohMauz}
E~G~D Cohen and D~Mauzerall.
\newblock A note on the {J}arzynski equality.
\newblock {\em J. Stat. Mech.}, page P07006, 2004.

\bibitem{mazonka}
O~Mazonka and C~Jarzynski.
\newblock Exactly solvable model illustrating far-from-equilibrium predictions.
\newblock 1999.
\newblock http://arXiv.org/cond-mat/9912121.

\bibitem{narayan}
O~Narayan and A~Dhar.
\newblock Reexamination of experimental tests of the fluctuation theorem.
\newblock {\em J. Phys. A}, 37:63, 2004.

\bibitem{EvansMS}
D~J Evans.
\newblock Relation between two proposed fluctuation theorems.
\newblock {\em Mol. Simulations}, 31:389, 2005.

\bibitem{dhar}
T~Mai and A~Dhar.
\newblock Nonequilibrium work fluctuations for oscillators in non-markovian
  baths.
\newblock {\em Phys. Rev. E}, 75:061101, 2007.

\bibitem{seifert05}
U~Seifert.
\newblock Entropy production along a stochastic trajectory and an integral
  fluctuation theorem.
\newblock {\em Phys. Rev. Lett.}, 95:040602, 2005.

\bibitem{seifert06}
T~Schmiedl, T~Speck, and U~Seifert.
\newblock Entropy production for mechanically or chemically driven
  biomolecules.
\newblock {\em J. Stat. Phys.}, 128, 2006.

\bibitem{imparatoa}
A~Imparato and L~Peliti.
\newblock Work-probability distribution in systems driven out of equilibrium.
\newblock {\em Phys. Rev. E}, 72:046114, 2005.

\bibitem{imparatob}
A~Imparato and L~Peliti.
\newblock Fluctuation relations for a driven {B}rownian particle.
\newblock {\em Phys. Rev. E}, 74:026106, 2006.

\bibitem{bsgj01a}
L~Bertini, A~De Sole, D~Gabrielli, G~Jona-Lasinio, and C~Landim.
\newblock Fluctuations in stationary nonequilibrium states of irreversible
  processes.
\newblock {\em Phys. Rev. Lett.}, 87:040601, 2001.

\bibitem{bsgj01b}
L~Bertini, A~De Sole, D~Gabrielli, G~Jona-Lasinio, and C~Landim.
\newblock Macroscopic fluctuation theory for stationary non-equilibrium states.
\newblock {\em J. Stat. Phys.}, 107:635, 2002.

\bibitem{lc97}
D~G Luchinsky and P~V~E McClintock.
\newblock Irreversibility of classical fluctuations studied in analogue
  electrical circuits.
\newblock {\em Nature}, 389:463, 1997.

\bibitem{GR04}
A~Gamba and L~Rondoni.
\newblock Current fluctuations in the nonequilibrium {L}orentz gas.
\newblock {\em Physica A}, 340:274, 2004.

\bibitem{GRV-jonaa}
C~Giberti, L~Rondoni, and C~Vernia.
\newblock Asymmetric fluctuation-relaxation paths in {FPU} models.
\newblock {\em Physica A}, 365:229, 2006.

\bibitem{GRV-jonab}
C~Giberti, L~Rondoni, and C~Vernia.
\newblock Temporal asymmetry of fluctuations in the nonequilibrium {FPU} model.
\newblock {\em Physica D}, 228:64, 2007.

\bibitem{PSR}
C~Paneni, D~J Searles, and L~Rondoni.
\newblock Temporal asymmetry of fluctuations in nonequilibrium states.
\newblock {\em J. Chem. Phys.}, 124:114109, 2006.

\bibitem{FW}
M~I Friedlin and A~D Wentzell.
\newblock {\em {Random Perturbations of Dynamical Systems}}.
\newblock (Berlin: Springer), 1984.

\bibitem{DeGroot}
S~R~De Groot and P~Mazur.
\newblock {\em {Non-Equilibrium Thermodynamics}}.
\newblock (Dover, New York), 1984.

\bibitem{HeatFlow}
D~J Searles and D~J Evans.
\newblock A fluctuation theorem for heat flow.
\newblock {\em Int. J. Thermophys.}, 22:123, 2001.

\bibitem{chatelier}
D~J Evans and D~J Searles.
\newblock Fluctuation theorem for {H}amiltonian systems: Le {C}hatelier{'}s
  principle.
\newblock {\em Phys. Rev. E}, 63:051105, 2001.

\bibitem{Denisov}
A~Denisov, H~M Castro-Beltran, and H~J Carmichael.
\newblock Time-asymmetric fluctuations of light and the breakdown of detailed
  balance.
\newblock {\em Phys. Rev. Lett.}, 88:243601, 2002.

\bibitem{ciliberto-1}
S~Ciliberto and C~Laroche.
\newblock An experimental test of the {Gallavotti-Cohen} fluctuation theorem.
\newblock {\em J. Physique IV}, 8:215, 1998.

\bibitem{feitosa}
K~Feitosa and N~Menon.
\newblock Fluidized granular medium as an instance of the fluctuation relation.
\newblock {\em Phys. Rev. Lett}, 92:164301, 2004.

\bibitem{shang}
X~D Shang, P~Tong, and K~Q Xia.
\newblock Test od steady-state fluctuation theorem in turbulent
  {Rayleigh-B\'enard} convection.
\newblock {\em Phys. Rev. E}, 72:015301, 2005.

\bibitem{tietz06}
T~Tietz, S~Schuler, S~Speck, U~Seifert, and J~Wrachtrup.
\newblock Measurement of stochastic entropy production.
\newblock {\em Phys. Rev. Lett.}, 97:050602, 2006.

\bibitem{ciliberto-2}
F~Douarche, S~Joubaud, N~B Garnier, A~Petrosyan, and S~Ciliberto.
\newblock Work fluctuation theorems for harmonic oscillators.
\newblock {\em Phys. Rev. Lett.}, 97:140603, 2006.

\bibitem{ciliberto-3}
F~Douarche, S~Ciliberto, A~Petrosyan, and I~Rabbiosi.
\newblock An experimental test of the {J}arzynski equality in a mechanical
  experiment.
\newblock {\em Europhys. Lett.}, 70:593, 2005.

\bibitem{blickle}
V~Blickle, T~Speck, L~Helden, U~Seifert, and C~Bechinger.
\newblock Thermodynamics of a colloidal particle in a time-dependent
  nonharmonic potential.
\newblock {\em Phys. Rev. Lett.}, 96:070603, 2006.

\bibitem{liphardt}
J~Liphardt, S~Dumont, S~B Smith, I~Tinoco, and C~Bustamante.
\newblock Equilibrium information from nonequilibrium measurements in an
  experimental test of {J}arzynski's equality.
\newblock {\em Science}, 296:1832, 2002.

\bibitem{ritort-1}
D~Collin, F~Ritort, C~Jarzynski, S~B Smith, I~Tinoco, and C~Bustamante.
\newblock Verification of the {C}rooks fluctuation theorem , recovery of {RNA}
  folding free energies.
\newblock {\em Nature}, 437:231, 2005.

\bibitem{trepagnier}
E~H Trepagnier, C~Jarzynski, F~Ritort, G~E Crooks, C~Bustamante, and
  J~Liphardt.
\newblock Experimental test of {Hatano, Sasa's} nonequilibrium steady-state
  equality.
\newblock {\em Proc. Nat. Acad. Sciences}, 101:15038, 2004.

\bibitem{ritort}
F~Ritort.
\newblock Work fluctuations, transient violations of the second law,
  free-energy recovery methods: Perspectives in theory, experiments.
\newblock {\em Poincar\'e Seminar}, 2:195, 2003.

\bibitem{PGaspnanotech}
P~Gaspard.
\newblock Out-of-equilibrium nanosystems.
\newblock {\em Progr. Theor. Phys. Suppl.}, 165:33, 2006.

\bibitem{Kantz}
E~G Altmann and H~Kantz.
\newblock Hypothesis of strong chaos and anomalous diffusion in coupled
  sympletic maps.
\newblock {\em Europhys. Lett.}, 78:10008, 2007.

\end{thebibliography}
\bibliographystyle{unsrt}

\end{document}